\PassOptionsToPackage{table,dvipsnames}{xcolor}
\documentclass[preprint,12pt]{elsarticle}

\usepackage{amssymb}
\usepackage{amsmath}

\journal{Journal of Systems and Software}

\usepackage[figuresright]{rotating}
\usepackage{xspace}
\usepackage{tikz}
\usepackage{pifont}

\usepackage{cite}
\usepackage{comment}
\usepackage{amsmath,amssymb,amsfonts}
\usepackage{algorithmic}
\usepackage{graphicx}
\usepackage{textcomp}
\usepackage{hyperref}
\usepackage{adjustbox}
\usepackage{fancybox,framed}
\usepackage{multirow,blindtext}
\usepackage{tabularx}
\usepackage{xurl}
\usepackage{ifthen}
\usepackage[utf8]{inputenc}
\usepackage{booktabs}
\newcommand{\mycomment}[1]{}
\usepackage{pdflscape}
\usepackage{cite}
\usepackage{amsmath,amssymb,amsfonts}
\usepackage{algorithmic}
\usepackage{graphicx}
\usepackage{textcomp}
\usepackage{amssymb}
\usepackage{siunitx}
\usepackage{pifont}
\usepackage{tcolorbox}
\usepackage[normalem]{ulem}
\usepackage{comment}
\usepackage{subcaption}
\usepackage{lscape}
\usepackage{listings}
\usepackage{xspace}
\usepackage{latexsym}
\usepackage{longtable} 

\usepackage{lipsum}  
\usepackage{tabularx}
\usepackage{quoting}

\lstdefinestyle{mystyle}{
    backgroundcolor=\color{backcolour},   
    commentstyle=\color{codegreen},
    keywordstyle=\color{codeblue},
    numberstyle=\tiny\color{codegray},
    stringstyle=\color{codegreen},
    basicstyle=\footnotesize\ttfamily,
    breakatwhitespace=false,
    breaklines=true,
    captionpos=b,
    keepspaces=true,
    numbers=left,
    numbersep=5pt,
    showspaces=false,
    showstringspaces=false,
    showtabs=false,
    tabsize=2
}

\definecolor{darkgreen}{rgb}{0,0.5,0}

\newcommand{\rev}[1]{\textcolor{black}{#1}}
\newcommand{\revTwo}[1]{\textcolor{black}{#1}}
\newcommand{\req}[1]{\textcolor{black}{#1}}
\newcommand{\revOne}[1]{\textcolor{black}{#1}}

\usepackage{listings}

\definecolor{codegray}{rgb}{0.5,0.5,0.5}
\definecolor{codegreen}{rgb}{0,0.6,0}
\definecolor{codeblue}{rgb}{0,0,0.6}
\definecolor{backcolour}{rgb}{0.95,0.95,0.92}

\usepackage{amssymb}
\usepackage{graphicx}

\definecolor{excerptgray}{HTML}{BEB8C6}
\definecolor{excerptblue}{HTML}{7AB1CE}
\definecolor{excerptgreen}{HTML}{33FFAC}

\begin{document}

\newcommand{\so}{Stack Overflow\xspace}
\newcommand{\cgpt}{ChatGPT\xspace}
\newcommand{\llama}{LLaMA\xspace}
\newcommand{\llamatwo}{\llama-2\xspace}

\newcommand{\dcircle}[1]{\ding{\numexpr171 + #1}}
\newcommand{\bcircle}[1]{\ding{\numexpr181 + #1}}

\newboolean{showcomments}
\setboolean{showcomments}{true}
\ifthenelse{\boolean{showcomments}}
 { \newcommand{\mynote}[2]{
      \fbox{\bfseries\sffamily\scriptsize#1}
        {\small$\blacktriangleright$\textsf{\emph{#2}}$\blacktriangleleft$}}}
        { \newcommand{\mynote}[2]{}}

\newcommand{\js}[1]{\mynote{Jordan}{\textcolor{red}{#1}}}
\newcommand{\lm}[1]{\mynote{Leuson}{\textcolor{blue}{#1}}} 
\newcommand{\Foutse}[1]{\mynote{Foutse}{\textcolor{purple}{#1}}}

\newcommand{\highlighttwo}[1]{%
\begin{tcolorbox}[leftrule=1mm,rightrule=1mm,toprule=0mm,bottomrule=0mm,left=0pt,right=0pt,top=0pt,bottom=0pt, colback=gray!30, colframe=gray!90]
#1
\end{tcolorbox}%
}

\begin{frontmatter}



\title{LLMs and Stack Overflow Discussions: Reliability, Impact, and Challenges}


\author[label1]{Leuson Da Silva}
\author[label2]{Jordan Samhi}
\author[label1]{Foutse Khomh}

\address[label1]{Polytechnique Montreal, Montreal, Canada}
\address[label2]{CISPA, Saarbrücken, Germany}

\begin{abstract}
Since its release in November 2022, \cgpt has shaken up \so, the premier platform for developers' queries on programming and software development.
Demonstrating an ability to generate instant, human-like responses to technical questions, \cgpt has ignited debates within the developer community about the evolving role of human-driven platforms in the age of generative AI.
Two months after \cgpt's release, Meta released its answer with its own Large Language Model (LLM) called \llama: \emph{the race was on}.
We conducted an empirical study analyzing questions from \so and using these LLMs to address them.
This way, we aim to 
\dcircle{1} quantify the reliability of LLMs' answers and their potential to replace \so in the long term;
\dcircle{2} identify and understand why LLMs fail; 
\dcircle{3} measure users' activity evolution with \so over time; and
\dcircle{4} compare LLMs together.
Our empirical results are unequivocal:
\emph{\cgpt and \llama challenge human expertise, yet do not outperform it for some domains}, while a significant decline in user posting activity has been observed.
Furthermore, we also discuss the impact of our findings regarding the usage and development of new LLMs \req{and provide guidelines for future challenges faced by users and researchers.} 
\end{abstract}




\end{frontmatter}



\section{Introduction}
\label{sec:introduction}
Practitioners must adopt different methods and approaches to support their work during software development. This involves making technical and non-technical decisions, encompassing the selection of programming languages, frameworks, and methodologies. These decisions have a prominent impact on the software development process~\citep{yli2016software, tamburri2013social,dias2020understanding}. 
When it comes to technical support, practitioners have become familiar with leveraging online Question and Answer (Q\&A) web forums as instrumental aids playing an essential role in different aspects, like API learning, code compression, and fixing or getting related problem information ~\citep{rubei2020postfinder, squire2015should}.
Notably, \so~\footnote{\url{https://stackoverflow.com/}} stands out as a preeminent online hub within the technology community~\citep{xia2017developers}.
However, owing to the inherent nature of these Q\&A platforms, which are contingent upon the accumulation of both individual and collective human knowledge, the potential for errors and the propagation of inaccurate information remains a concern~\citep{verdi2020empirical, ragkhitwetsagul2019toxic, zhang2018code}.

With the advance of Large Language Models (LLMs), practitioners and researchers have explored the potential of leveraging LLMs for different software engineering tasks; \req{for example, pair programming assistants and simulating human behavior as questionnaire respondents} ~\citep{dakhel2023github, hamalainen2023evaluating, liang2023can, hou2023large}.
\req{Such a variety of usage is motivated by the capability of LLMs to provide comprehensive and concise explanations for diverse subjects instantaneously}~\citep{touvron2023llama}.
Furthermore, with the release of \cgpt\footnote{\url{https://openai.com/chatgpt}}, publicly available for the general audience, users have prompted LLMs for different purposes, like simulating a Q\&A online web forum.
As a result, these models, by engaging in human-like tasks, can continuously learn from user feedback (e.g., via fine-tuning) and improve their accuracy~\citep{bakker2022fine}.

Therefore, practitioners might turn to using more and recurrently LLMs instead of checking truthful and verifiable sources.
Nonetheless, exclusively relying on information generated by LLMs is a threat, as LLMs might generate inaccurate information~\citep{goodrich2019assessing}.
To overcome the imminent risks, \so has proactively announced prohibiting the use of information generated by LLMs~\citep{stack2023ai-ban}.
However, considering the difficulty of detecting contents generated by LLMs~\citep{tang2023science, sadasivan2023can}, they later announced OverflowAI, an integration of generative AI into the platform impacting users, products, and IDEs~\citep{stack2023overai}.
Even under these new circumstances, it is still necessary to assess the generated answers and determine the extent to which LLMs can be effectively utilized.  

\rev{Such concern is becoming a general and recurrent topic, also targeted by other research areas.
For example, Lee et al. \citep{lee2023benefits} investigate the reliability of content generated by \cgpt-3.5 in medicine.
Based on scenario examples of potential medical use, the authors evaluate the current abilities of \cgpt under different situations.
The model performs satisfactorily, though the authors also observe some hallucinations, like inferring information unrelated to previously shared content.
These hallucinations were further explored in \cgpt 4 when the model could catch them.}

\rev{In the same way, evaluating the reliability of LLMs in Software Engineering while assisting developers with their daily tasks is a relevant and important topic, bringing the attention of previous studies.}
Kabir et al. \citep{kabir2023answers} and Liu et al. \citep{liu2023better} investigate the correctness of \cgpt by comparing the answers reported with human ones.
\revOne{However, by reliability, we refer to the ability of LLMs to consistently produce correct and accurate behavior across time and varying conditions.
While correctness is a key aspect of evaluating LLMs, reliability encompasses more than just factual accuracy. LLMs might provide correct answers, but they could also share content that, although technically correct, may still be misleading or harmful to users in some contexts. 
Therefore, evaluating such an aspect is crucial for ensuring LLMs do not produce misleading, harmful, or incomplete information, even when their responses are technically correct.
\revTwo{Furthermore, previous studies focus only on \cgpt, while they also do not deeply explore the most challenging domains faced by the model and how users deal with them.}
\req{Given LLMs' current diversity, different number of parameters supported by models, and features, it is important to consider these factors when investigating and comparing findings across different models.}
}

\revTwo{Researchers have also investigated the impact of LLMs in \so. 
Even before the release of LLMs, a decline in the number of answers to 
questions related to different programming languages was observed on \so \citep{blanco2020understanding, syam2023empirical,pragmatic2023stackoverflow}. 
Recently, Rio-Chanona et al. \citep{del2023large} report an overall decrease in posting activity on \so after the release of \cgpt.
However, whether such a decrease is uniform in different domains is unknown, as different topics involve different groups and their engagement.
The decrease in users' activity on \so not only affects how developers receive support for their issues but also influences future versions of LLMs, given the need to train new models with updated data. 
}

Aiming to address the previous gaps, in this paper, we explore LLMs' reliability, challenges, and impact on \so, focusing on exploring different models and users' activity regarding different topics and groups. 
To that end, we perform an empirical study analyzing questions and associated answers mined from \so.
First, we mine questions and related information from \so (before and after \cgpt's release).
\req{Then, we analyze the overall and by topics impact of posting activity.}
Second, following previous studies conducted on \so, we select a representative number of the previously selected questions to prompt LLMs, further comparing the generated answers with the original ones provided on \so.
\rev{To mitigate previous studies' diversity issues regarding LLMs under analysis, we investigate two models: \cgpt-3.5 and \llama-2-7b.}
\req{Next, based on the answers provided, we further manually analyze the ones that LLMs present low similarity compared with human answers, reporting the most challenging topics they faced.}

We observe that \cgpt-3.5 significantly outperforms \llamatwo-7b regarding answers' textual similarity.
Although \llama-2-7b does not present significantly better results than \cgpt, this model represents an alternative, free, and public option for the tech community.
About the challenges faced by the LLMs, overall, they performed well in most domains, with some challenges regarding questions targeting \textit{Frameworks and Libraries}, leading to different observed impacts on specific topics.
Although \cgpt presents a more neutral sentiment when generating answers, 
\llama presents a more positive sentiment.
\req{Finally, regarding the impact on \so, our results confirm previous findings about a continuous significant decline in posting, answering, and commenting activity after \cgpt's release. 
Such a decline is not uniform initially, as we observed no statistical decline for some topics, like specific frameworks and libraries. 
However, after one year of \cgpt's release, we observe a constant and significant drop in posted content in \so.}

Our paper makes the following contributions:
\begin{itemize}
    \item We report an empirical study evaluating the reliability of answers provided by \cgpt and \llama in comparison to human-generated answers on \so.
    \item \rev{We identify and further explore the domains where LLMs fall short and show that for some topics, like frameworks and libraries.}
    \item \rev{We compare \cgpt with \llama and show that despite the structural difference between these models, they share similar challenges. }
    \item We reinforce previous findings about a decline in posting activity on \so, though our results also show that such a decline is not uniform on different topics and groups. 
    \item \rev{We provide a dataset of questions from \so and associated scripts available through our online Appendix \citep{appendix} and Zenodo as well.\footnote{\url{https://doi.org/10.5281/zenodo.15086541}}}
\end{itemize}

\rev{The rest of the paper is organized as follows: Section~\ref{sec:background} presents details about the LLMs evaluated here, while Section~\ref{sec:study-setup} explains our study setup and the steps performed. 
In Section~\ref{sec:results}, we present the results, which are further discussed in Section~\ref{sec:discussion}. 
Threats to the validity of our study and related work are discussed in Sections~\ref{sec:threats} and ~\ref{sec:related-work} respectively.
Finally, in Section~\ref{sec:conclusion}, we present our conclusions.}

\section{Background}
\label{sec:background}
This section introduces the necessary concepts and terminology used in our study.

\noindent
\textbf{\so}:
\so (SO) is a platform where developers can ask questions and share their knowledge with others \citep{barua2014developers}.
It is designed to provide accurate answers to specific programming problems, fostering a community of learning and sharing among developers.
Over time, developers have adopted this platform to discuss different topics, while building general knowledge with the tech community.

\noindent
\textbf{Large Language Model}:
A large language model (LLM) is a machine learning model trained on extensive sets of types of data, like textual, images, audio, and structured data.
Recently, LLMs trained on textual data brought the attention of users, scientists, and maintainers due to their support of generating human-like text \citep{xu2022systematic}.
With millions to billions of parameters, LLMs are used for various natural language processing tasks.
Different LLMs have been proposed for tasks related to software engineering, like StarCoder \citep{li2023starcoder}, Copilot \citep{copilot}, and CodeBERT \citep{feng2020codebert}.

\noindent
\textbf{\cgpt}:
\cgpt is a language model developed by OpenAI. 
It is designed to generate human-like text based on the prompts it receives. 
Built on the GPT architecture, it can be used for a variety of applications, among which are code generation, summarization, translation, testing, and documentation \citep{zheng2023towards, ozkaya2023application}.
Currently, OpenAI offers different models, but in this study, we focus on \cgpt-3.5.
\rev{The motivation for this decision is driven by the recurrent adoption of this version by related work; furthermore, it is the default version that users have access to through the app option without further costs.}

\noindent
\textbf{\llama}:
\llama, released by Meta AI, is a series of large language models that come with different sizes (e.g., 7, 13, and 70 billion parameters for \llama-2) \citep{touvron2023llama}.
Similar to \cgpt, it spans various use cases, such as interacting with humans. 
In our study, we relied on the latest \llama version, i.e., \llamatwo.
\rev{This version has similarities with \cgpt, like its architecture based on transformers and its generative capabilities, leading to generate coherent and contextually relevant text based on input prompts.}
Although this model has not been explored like \cgpt, we aim to investigate how these models differ and complement each other.
\rev{Specifically for this study, we consider  \llama-2-7b as it represents the most accessible option for users, based on the required resources to load and use the model locally.}

\req{Although there are newer versions of the selected LLMs in this study, like GPT-4 and LLaMA 4, not evaluated here, 
we consider the selected versions based on their popularity and availability for the general users. 
\revOne{Our study was conducted in two phases: first, we analyzed data before and after the release of ChatGPT, then, one year later, we collected data for the one-year milestone. 
When we conducted our study for the first phase in May 2023, GPT-3.5-turbo was the publicly available model. 
GPT-4 was released in March 2023 but was only accessible to (i) ChatGPT Plus subscribers or (ii) API users with a waitlist. This way, our analysis was necessarily based on \cgpt-3.5, the model available to the general public at that time through its GUI without additional costs compared with \cgpt-4.}
Regarding \llama, newer versions are known for handling more parameters and, consequently, dealing with more complex tasks.
\llama API shares the same costs previously discussed, though their models can be locally loaded by users in their environments. However, it requires advanced resources due to hardware and software resource constraints, which might be challenging for some users.
}

\section{Study Setup}
\label{sec:study-setup}

Our experimental methodology investigates the reliability, challenges, and impact of LLMs on \so.
To this end, our study aims to answer the following research questions:

\begin{itemize}
    \item \textbf{RQ1.} \emph{How does the reliability of answers generated by \cgpt and \llamatwo compare to those provided by \so users?} 

\req{Previous studies have investigated the adoption of LLMs to address \so questions by evaluating the potential of \cgpt for such a task \citep{kabir2023answers, liu2023better}.
Knowing the diversity of LLMs available and their different configurations, we take a different route by evaluating the reliability of different LLMs while comparing them.
Reliability in Software can be understood as the probability of having failure-free software execution for a timely and spacely specified period \citep{ieee1990standard}.
In this study, we investigate reliability based on the capability of LLMs to report answers matching the specifications reported in questions. 
Knowing that different answers can fix a single question in \so, we focus on the answer previously accepted by the requester.
\revOne{To evaluate the reliability of LLM-generated answers, we use the accepted answer on \so as a proxy, since it is typically considered the most reliable solution by the community, based on user feedback.}
We aim to investigate the capability of LLMs to report reliable and valuable content to users, considering the occurrence of wrong answers due to the LLM faults \citep{lyu2007software}.}
This way, we can evaluate how far or close LLMs are to the most accurate information that properly fixes the problem in that specific context.
Following previous studies \citep{yazdaninia2021characterization, asaduzzaman2013answering}, we randomly select a representative number of previously collected questions and prompt the models to answer them.
Next, we compute the reliability by checking the textual and semantic similarity of the generated and original answers.
\req{Finally, knowing that answers avoiding negative comments are more likely to be accepted \rev{by \so users}~\citep{calefato2015mining}, we aim to assess how close the generated answers fit with general users' preferences. 
For that, we perform a sentiment analysis of the generated answers.}
    
    \item \textbf{RQ2.} \emph{What specific domains pose challenges for \cgpt and \llamatwo when generating reliable and accurate answers?} 
\req{When evaluating the adoption of LLMs to address \so questions, previous studies' findings report how well LLMs perform \citep{kabir2023answers, liu2023better}.
However, there is a gap regarding how LLMs perform in different domains.
Aiming to explore such a perspective, in this RQ, we investigate how to leverage LLMs' adoption for practitioners.
Based on the results of RQ1, for a subset of questions (low and high text similarity), we analyze and classify their associated domains, based on the categorization proposed by Barua et al. \citep{barua2014developers}, reporting the domains that LLMs performed well and the challenging ones.
}
Second, we explore the possible structural factors of the questions that correlate with high or low similarity.

    \item \textbf{RQ3.} \emph{\revOne{What is the impact of \cgpt’s public release on \so posting activity?}}

        \begin{itemize}
            \item \rev{\textbf{RQ3.1.} \emph{How are different domains from \so impacted by the release of \cgpt?}}
        \end{itemize}
\req{Rio-Chanona et al. \citep{del2023large} and Burtch \citep{burtch2023consequences} investigate the impact of \cgpt in \so. 
Although they report a decline in user posting activities and \so visits, it is unknown whether such a decline is uniform for different domains, like programming languages, frameworks, and libraries.
Furthermore, they investigate the immediate impact without checking the evolution over time.
Burtch et al. \citep{burtch2023consequences} even provide an analysis by individual topics but do not group related ones and analyze them together.
}
Aiming to bring light to these concerns, we first mine questions and related information from \so for five months before and after the release of \cgpt (30$^{th}$ of November 2022), and after one year of the release (30$^{th}$ of November 2023). 
Second, we statistically assess the impact on the usage/engagement by comparing the number of posted questions, answers, comments, and related information.
Third, to investigate the impact of \cgpt on specific domains of \so, we filter previously mined questions, which represent challenges for LLMs, and compare the frequency of posted questions referring to the associated tags.
Based on our findings, we discuss possible factors that might influence the results and the different strategies for adopting/combining \so and LLMs from now on.

\item \rev{\textbf{RQ4.} \revOne{\emph{How has the release of \cgpt impacted the activity patterns
of \so users?}} 
\req{When investigating the impact of \cgpt in \so, Rio-Chanona et al. \citep{del2023large} mostly focus on the posting activity, which does not cover how such an impact affects the activity of users in \so.
Burtch et al. \citep{burtch2023consequences} and Xue et al. \citep{xue2023can} even investigate the impact on users, focusing on the access traffic and the way users deal with \so, respectively. 
In this RQ, we take a different route by investigating the impact of active users in \so.}
First, we compare users' activity level on \so pre- and post-release of \cgpt, considering posting questions, answers, and comments. 
Second, based on the different groups of users observed (reminiscent and new users), we further explore how users combine the adoption of \cgpt and \so. 
For that, we manually analyze questions not initially addressed by the LLM and later posted on the web forum.
Based on our findings, we discuss the challenges \so faces regarding their new users while understanding their new current needs.}
\end{itemize}

\begin{figure*}[t]
    \centering
    \includegraphics[scale=0.24]{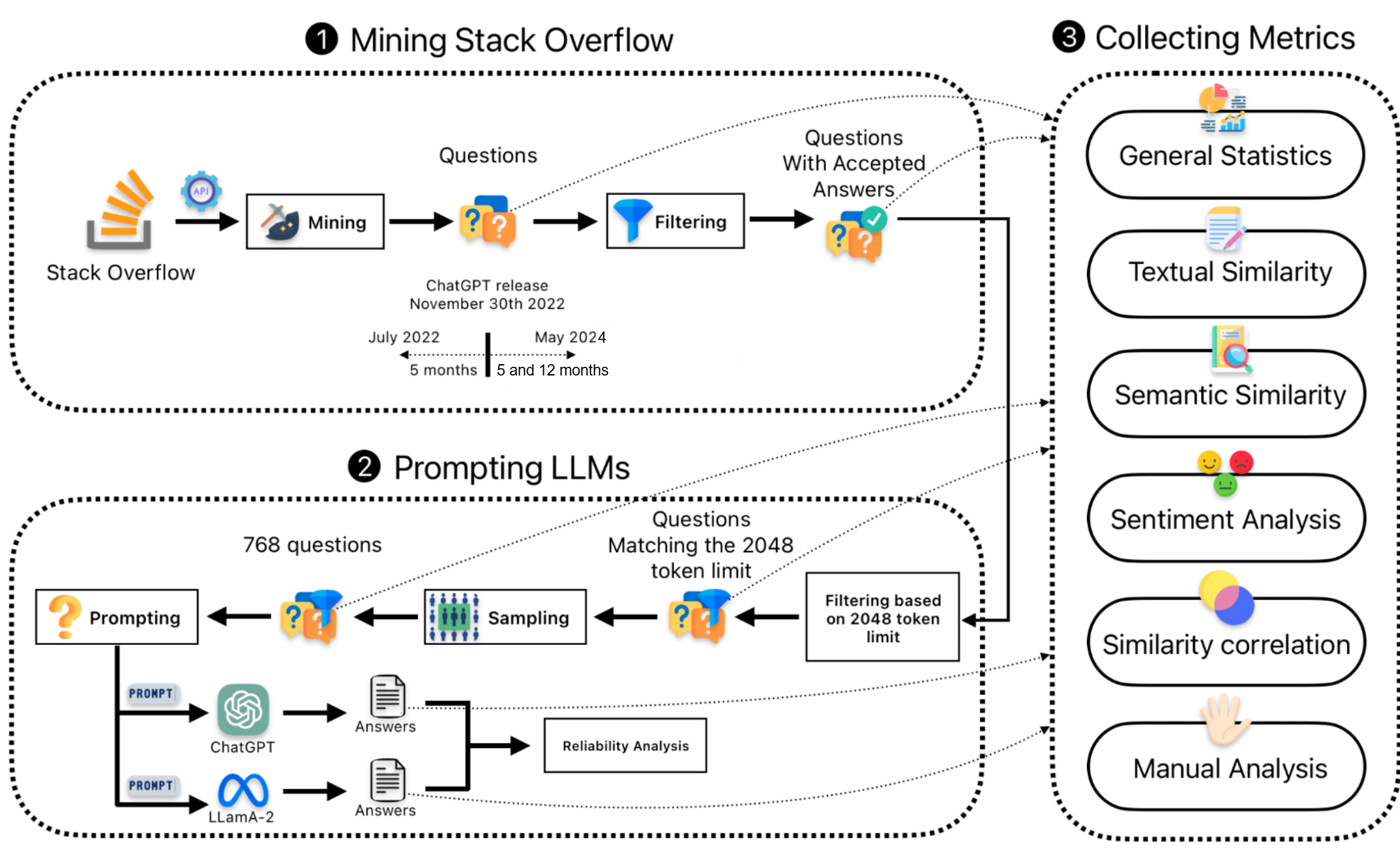}%
    \caption{Overview of our empirical setup}
    \label{fig:overview}
\end{figure*}

The remainder of this section describes the empirical setup of our study.
We structure our study in three main steps, as presented in Figure~\ref{fig:overview}:
\bcircle{1} we mine SO questions using its API from July 2022 till May 2023, and Nov 2023 till May 2024;
then, we filter the questions with accepted answers and sort them based on their score.
\bcircle{2} we select a representative number of questions from this subgroup and use them to prompt \cgpt and \llamatwo; and
\bcircle{3} we further collect metrics regarding question contents and associated information.

\subsection{Mining SO Data}
\label{sec:mining}

To investigate the current impact of \cgpt on \so usage, \req{following previous studies' methodological steps \citep{del2023large}}, we mined SO posts and collected the data using SO's API\footnote{\url{https://data.stackexchange.com/stackoverflow/query/new}} from 5 months before and after the \cgpt release \rev{(1$^{st}$ July till 29$^{th}$ November 2022, and 30$^{th}$ November 2022 till 30$^{th}$ April 2023, respectively)}.
\req{Knowing that adopting the release date of \cgpt as a cutoff may introduce bias in our results, we also mine posts dated one year after the release (30$^{th}$ November 2023 till 30$^{th}$ April 2024, respectively). This way, we can evaluate the long-term impact, considering immediate reactions and the more sustained engagement over time. 
This approach ensures a more comprehensive analysis, capturing shifts in discussions, challenges, or resolutions that might emerge only after extended use or broader adoption. 
From now on, when we say data after \cgpt's release, we mean the data posted just after the release.
To refer to the data posted after one year of the release, we say one-year milestone.}
\so's API limits the returned data for a single request to \num{50000} rows.
Hence, we performed as many requests as needed to gather all relevant data.\footnote{\so offers its archived dump dataset quarterly. By the time of our mining, we needed up-to-date information, leading us to use the provided API.}
\rev{For example, to get the \texttt{posts} posted between November 29-30th, we used the following query:}

\begin{tcolorbox}[colback=white,colframe=black]
\texttt{select * from Posts where Posts.CreationDate \textgreater `2022-11-28 00:00:00' and Posts.CreationDate \textless `2022-11-30 00:00:00'}
\end{tcolorbox}

\rev{
This way, we make sure the contents posted were created before or after the release of \cgpt.}

\begin{table}
\centering
\begin{tabular}{lrrr}
\hline \hline
& \multicolumn{1}{c}{\begin{tabular}[c]{@{}c@{}}Pre-Release\end{tabular}} & \multicolumn{1}{c}{\begin{tabular}[c]{@{}c@{}}Post-Release\end{tabular}} & 
\multicolumn{1}{c}{\begin{tabular}[c]{@{}c@{}}One-Year Milestone\end{tabular}} \\ \hline
\hline
Questions & \num{674425} & \num{578115} & \num{296718} \\ \hline
Questions with Answers & \num{424808} & \num{329412}  & \num{167303} \\ \hline
\begin{tabular}[c]{@{}l@{}}Question with Accepted Answers\end{tabular} & \num{212775} & \num{157187}  & \num{78262} \\ \hline
Questions without Answers & \num{249617} & \num{248703}  & \num{129415} \\ \hline
Answers & \num{543533} & \num{425391}  & \num{202326} \\ \hline
Comments & \num{1673906} & \num{1348982}  & \num{718496} \\ \hline
Tags & \num{36837} & \num{37328}  & \num{31867}
\end{tabular}
\caption{Sample data description}
\label{tab:mining}
\end{table}

Initially, we collected the posts and their related comments (as returned by \so' API).
However, in \so's data, questions and answers are interchangeably treated as regular posts, differentiating from each other based on their type.
Furthermore, while questions can receive multiple answers, only one can be targeted as the accepted one; posts can be associated with multiple comments.
We collected all information available for each element under investigation, including the creation and last modification date, title, description, score, views, etc. 
The creation and modification dates are essential information since a question might have been added before the \cgpt release, but its accepted answer might have been dated after \cgpt's release. 
Next, we organize the data by associating the questions with their related answers and comments.
This step is required since answers and comments for questions not dated from the interval evaluated here might introduce bias in our results. 
\req{For example, a question posted before the release of \cgpt may be answered by a user that used the LLM to address the initial reported issue.}

This way, we adopt one particular constraint to investigate the overall impact of posting activity on \so. 
\rev{For the questions posted before \cgpt release, we only consider valid answers and comments, the ones also dated before the release date (30th November 2022).}
\req{Finally, we were able to collect \num{674425}, \num{578115} and \num{296718} questions before, after, and one-year milestone \cgpt's release, respectively. 
Table \ref{tab:mining} presents an overview of our dataset.
Overall, we notice a drop of almost \num{100000} between the before and after the release of \cgpt (for the same amount of time, i.e., five months). 
One year after its release, the drop becomes even more pronounced, decreasing by almost threefold (approximately \num{300000} questions).}

\rev{On the other hand, to investigate the impact of questions addressing specific topics (programming languages, frameworks, and libraries), we first properly filtered the questions focusing on these topics.
When users ask about these topics, they commonly add the name of the target topic as one of the question tags; for example, \texttt{Java}, \texttt{Angular}, and \texttt{Pandas}.
Since multiple tags can be used for a single question, we counted the number of times a single tag was adopted.
Then, we organize the frequency of questions associated with each topic daily. 
Although users may consider variations or abbreviations when targeting their subjects, we observe a consistent adoption of the official topic names; \req{for example, \texttt{Pandas} can be abbreviated to \texttt{pd}.}
}

\rev{Finally, to investigate the impact on the user's activity patterns in \so, we evaluate the user information associated with the final set of questions, answers, and comments collected previously. 
For that, we group the users associated with each content into three groups: \emph{questioners}, \emph{respondents}, and \emph{commentators}.
This way, we could compare the number of users involved in each type of activity in \so.
To answer RQ4, we collected further user information, including their account creation date, allowing us to identify \textit{reminiscent} and \textit{new} users after the release.
We request each user's information using their IDs on the \so API.
} 

\subsection{Prompting LLM Models}
\label{sec:prompting-llm}
To investigate LLMs' capabilities to generate high-quality answers compared to humans' answers, we selected questions from our dataset with \emph{accepted answers only} (refer to Table \ref{tab:mining}).
To adhere to the 2048 token limitation for the LLMs under study, questions exceeding this token count were excluded (\num{6998} questions spanning before \cgpt's release were filtered, and \num{11659} after \cgpt's release).
Token requirements were calculated based on the question title, description, and associated tags\footnote{\url{https://pypi.org/project/tiktoken/}}.
As a result, the dataset comprised \num{205777} questions before (i.e., \num{212775} - \num{6998}) and \num{145528} questions after the release of \cgpt (i.e., \num{157187} - \num{11659}).
For subsequent analysis, we randomly selected 384 questions from each set (\rev{before and after \cgpt release}), aiming for a 95\% confidence level and a 5\% margin of error.

\begin{figure}
    \centering
    \includegraphics[width=0.65\linewidth]{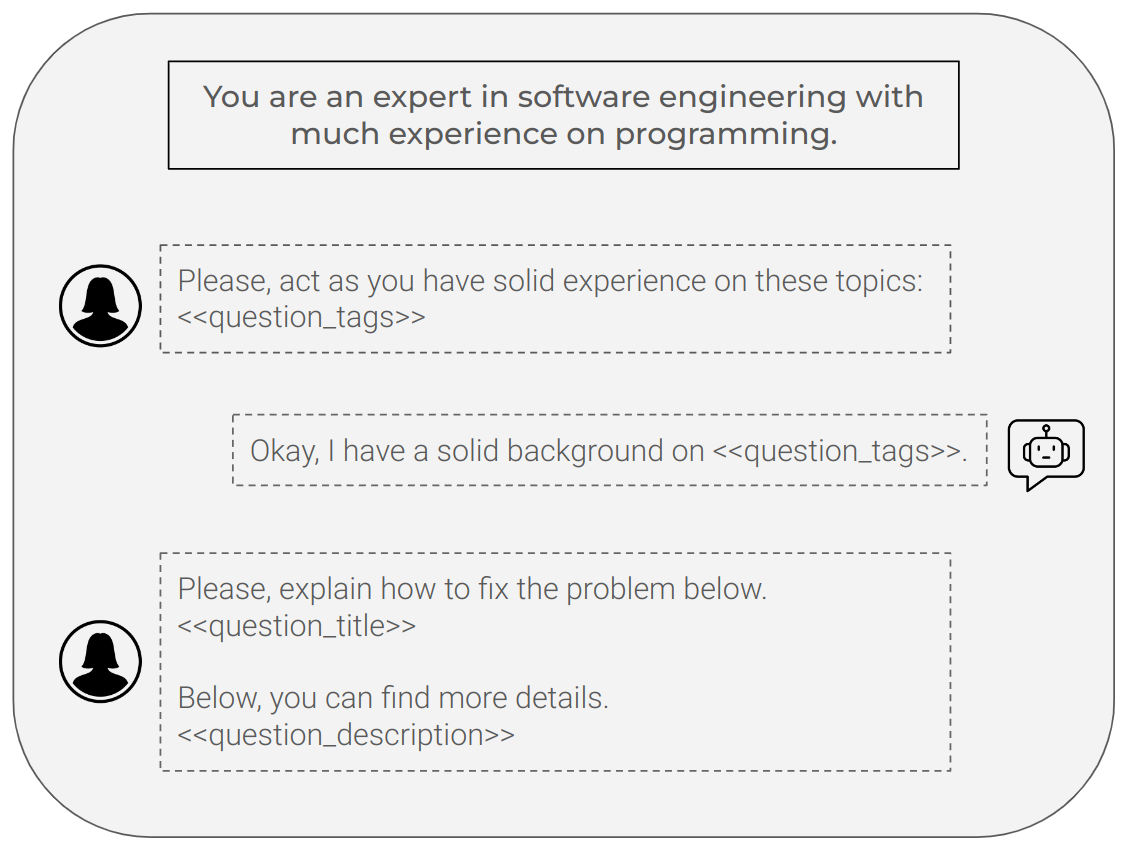}
    \caption{Prompt approach adopted to prompt the LLMs. Question contents were given as input, replacing the elements informed highlighted with \textless{}\textless{} and \textgreater{}\textgreater{}, respectively.}
    \label{fig:prompt-approach}
\end{figure}

This study focuses on two LLMs: \cgpt 3.5 and \llamatwo-7b. 
We used the GPT-3.5-turbo variant for our research, interacting with it through its API. 
Unlike \cgpt, we call \llama by locally loading it.
For our experiments, \revOne{we used the 7B-sized \llamatwo model (\texttt{Llama-2-7b-chat-hf} checkpoint in its base configuration), which is the 7-billion-parameter chat-optimized variant, ensuring alignment with conversational tasks.} 
The model was loaded via Hugging Face support\footnote{\url{https://huggingface.co/meta-llama/Llama-2-7b}}, on a machine equipped with 40GB RAM.
When prompting the LLMs, we follow a standardized approach that simulates a prior conversation to establish context.
For that, we consider the tags associated with the questions as the foundation for instructing the LLM to adopt the persona of an expert with extensive knowledge in the relevant subject matter\rev{\citep{white2023prompt}}.
Next, we consider the question's title to properly ask the LLM to explain how to fix the problem.
Additionally, we argue that further context will be given by informing the question description.
\rev{Figure \ref{fig:prompt-approach} presents the adopted approach with more details.}
Regarding the usage of these models, we consider their default configurations. 
\revOne{For example, we prompt the models with their default temperature and context length (maximum token limit) \citep{wagner2024towards}.
By making this decision, we aim to avoid arbitrary bias, considering the diversity of tasks addressed in this study. Here, we focus on evaluating the models in similar ways users would, such as respecting their default configurations and interacting with them under typical usage scenarios without additional hyperparameter adjustments.
Furthermore, related studies did not report specific hyperparameters, leading us to believe they used default configurations as well.} 
Finally, for each question, we prompt each model once.

\subsection{Research Questions: Evaluation Design}
\label{sec:collecting-metrics}
This section presents the methodology adopted to address each RQ and the associated metrics adopted when running our study.

\subsubsection{RQ1: Comparing Human and LLM-Generated Answers for \so Questions}

\textbf{Textual Similarity:}
To assess the similarity between answers from \so and those generated by LLMs, we employed the cosine similarity metric ($\theta$) ~\citep{salton1988term,lahitani2016cosine}.
To compute this metric, we relied on a pre-trained Sentence Transformers model (\texttt{all-MiniLM-L6-v2})\footnote{https://github.com/UKPLab/sentence-transformers}.
First, we take the accepted and LLM-generated answers for a given question and compute their embeddings.
Second, we used the function \texttt{cos\_sim} from the same model, passing the previous embeddings and computing the metric.

\textbf{Semantic Similarity:}
Knowing that the cosine similarity is sensitive and does not compute the semantics between two given answers, we might misclassify similar answers due to low textual similarity. 
Motivated by previous studies \citep{liang2023can}, we decided to compute the semantic similarity relying on the judgment capability of the LLMs evaluated in this study. 
\revOne{Our goal is not to evaluate the reliability of a particular semantic similarity metric, but rather to analyze how the LLMs themselves assess semantic similarity between answers. Specifically, we investigate the degree of consistency in their judgments and how closely they align with human evaluation.}
First, we simulate a prior conversation to establish the context, using the questions' tags again to instruct the LLM to adopt the persona of an expert.
Second, we ask the LLM to compare the original and LLM-generated answers, reporting an evaluation using a VERY LOW to VERY HIGH scale.
By combining both answers in one single prompt, we reached the maximum number of tokens by the LLMs; \cgpt did not classify one answer, while \llama did not classify 72 answers due to resource constraints, resulting in 1,464 answers that could be analyzed.
Third, to evaluate the consistency and level of agreement among different runs for the same model, we prompted ChatGPT to classify the 1,535 answers three times.
\req{In the end, we randomly selected a subsample of 91 questions from each group (ensuring a 95\% confidence level with a 10\% margin of error).
} \req{The primary author manually labeled these questions, following the same scale previously adopted by the LLMs, by comparing the contents of the human-reported and LLM-generated answers.
\revOne{Since the categories adopted present an ordinal structure, we must consider such aspect when assessing the level of inter-rater agreement between our manual analysis and the assessments provided by LLMs, and the agreements between the LLMs themselves.
As categories have a defined order, the distance between them may not be equal; for example, disagreeing between VERY HIGH and HIGH is less severe than between VERY HIGH and VERY LOW.
As a result, we compute Krippendorff's alpha coefficient ~\citep{krippendorff2011computing} with an ordinal distance function, as this coefficient handles ordinal weighting by default.}
}

\textbf{Sentiment Analysis.}
In light of research suggesting that avoiding negative comments can improve answer acceptance on Stack Overflow~\citep{calefato2015mining}, we aim to investigate whether LLMs' responses align with this observation.
To do so, we perform a sentiment analysis on answers generated by the LLMs using a pre-trained Transformers model (\texttt{sentiment-analysis})\footnote{https://github.com/huggingface/transformers}.
Subsequently, we statistically compared the sentiment outcome achieved by each LLM using the Wilcoxon test.

\subsubsection{RQ2: Extracting Meta Information and Associated Domains from Questions}
\label{sec:manual-low-high}
To answer this RQ, we perform a manual analysis focusing on the questions linked to answers reporting low and high textual similarity for each LLM evaluated.
For that, we select the questions associated with the first and fourth quartile of each LLM violin plot (low and high similarity, respectively; \revOne{we explain in detail when discussing our results in Section \ref{sec:results}}). 
\rev{Since the main focus here is to report the domains that challenge LLMs,} \revOne{we adopt the taxonomy provided by Barua et al. \citep{barua2014developers} as a starting point for our categorization. 
When analyzing the proposed categories, we observe that they were too specific, limiting their application to our sample. 
This way, we decided to group similar categories into single ones.
For example, Barua et al. \citep{barua2014developers} report \textit{UI Development},  \textit{Website Design/CSS}, \textit{Web Development}, and  \textit{Web Service/Application} as individual categories. 
For our study, we group all of them into \textit{Web Development}, as they share a similar goal. 
As a result, we define nine categories: \textit{Database/SQL}, \textit{Framework/Library}, \textit{General Programming Concepts}, \textit{Mobile}, 
\textit{Networking}, \textit{Operating Systems}, 
\textit{Programming Languages}, \textit{Tools/IDEs}, and \textit{Web Development.}
However, when performing the analysis, we observe that some questions could not be classified using the previous categories.
By analysing them, we additionally report two categories: \textit{DevOps} and \textit{ML/Datascience}.
Finally, our analysis is based on 11 categories, nine derived from previous work and two new ones reported by us. 
}

\req{Aiming to get more understanding of these questions, we decided to further explore them based on the meta information we could extract. 
Initially, we explored some questions to identify the information shared among them. \revOne{We adopted a simple process, by randomly selecting questions previously prompted to LLMs (no more than 20 questions).
As our main goal was to specify the types of information shared by users when posting their questions, when we observed that no new type of information was reported (saturation), we stopped the exploration.}
As a result, we defined a spreadsheet, incorporating the information we could extract and their predefined values, providing a guide for the manual analysis.
Table \ref{tab:aspects-so} provides the information we extracted. 
For this analysis, we focus only on the aspects associated with the questions.
}

\textbf{Similarity Correlation.}
To explore possible factors contributing to the higher reliability of answers generated by LLMs, we extended our analysis to include the context in which the models were prompted.
\revOne{Overall, we performed a correlation analysis between (i) the textual similarity achieved by a given LLM answer, previously computed, and (ii) the number of words associated with the different types of context we have.}
\revOne{Since we have three different types of context in our study (tags, title, and description), we adopted distinct strategies for each, as reported below.
For tags, we computed the number of tags associated with each question, as this directly reflects the context provided by the tags.
For both the title and description, we first computed the number of words. Then, we split them into intervals: a 2-word interval for the title and a 25-word interval for the description. 
The rationale behind this segmentation was to capture the varying levels of context provided by different text lengths.
For titles, a 2-word interval was chosen because we did not observe any questions with only a single word. 
For descriptions, we opted for 25-word intervals, as there was a consistent rise in the number of questions at each interval length (e.g., 25, 50, etc.).
Our goal with this analysis was to quantify the impact of context size on model performance, revealing how the amount of input text influences the quality of the LLM's responses.}
Since our dataset was not drawn from a normally distributed population, as confirmed by a \textit{Shapiro-Wilk} test, we used the non-parametric Spearman test.
Initially, we verify whether the context used to prompt the LLMs correlates with the textual similarity level previously computed (for that, we rely on the Spearman test).

\newpage

\begin{landscape}

\begin{table}[]
\begin{tabular}{clllc}
\hline
\hline
                          & \multicolumn{1}{c}{Structural Aspects} & \multicolumn{1}{c}{Definition}                                                                                 & \multicolumn{1}{c}{Associated Values}                                                     & Target RQs  \\ \hline
                          \hline
\multirow{14}{*}{Question} & Context Adequacy                       & \begin{tabular}[c]{@{}l@{}}The level of information \\ provided in the question.\end{tabular}                  & Shallow, Medium, Strong                                                                    & \multirow{14}{*}{RQ2, RQ4} \\ \cline{2-4}
                          & External Reference                     & \begin{tabular}[c]{@{}l@{}}External content linked \\ with the question\end{tabular}                           & \begin{tabular}[c]{@{}l@{}}Another SO question, \\ documentation, link\end{tabular}       &                           \\ \cline{2-4}
                          & No Textual Content                     & \begin{tabular}[c]{@{}l@{}}Adoption of content \\ different than text\end{tabular}                             & Image                                                                                     &                           \\ \cline{2-4}
                          & Illustrative Context                   & \begin{tabular}[c]{@{}l@{}}Supportive information \\ provided to illustrate the \\ target problem\end{tabular} & \begin{tabular}[c]{@{}l@{}}Code, Error Message, \\ Outcome, Expected Outcome\end{tabular} &                           \\ \cline{2-4}
                          & Goal                                   & \begin{tabular}[c]{@{}l@{}}The approach adopted \\ to ask for assistance\end{tabular}                          & \begin{tabular}[c]{@{}l@{}}Question, Explanation, \\ Open\end{tabular}                    &                           \\ \cline{2-4}
                          & Domain                                 & \begin{tabular}[c]{@{}l@{}}The domain associated\\ with the question\end{tabular}                              & Categories adapted from \citep{barua2014developers}                                                                  &                           \\ \cline{2-4}
                          & Tag Concordance                        & \begin{tabular}[c]{@{}l@{}}The level of concordance \\ among the informed tags\end{tabular}                    & Shallow, Medium, Strong                                                                   &                           \\ \hline
\multirow{6}{*}{Answer}   & No Textual Content                     & \begin{tabular}[c]{@{}l@{}}Adoption of content \\ different than text\end{tabular}                             & Image                                                                                     & \multirow{6}{*}{RQ4}      \\ \cline{2-4}
                          & External Reference                     & \begin{tabular}[c]{@{}l@{}}External content linked \\ with the answer\end{tabular}                             & \begin{tabular}[c]{@{}l@{}}Another SO question, \\ documentation, link\end{tabular}       &                           \\ \cline{2-4}
                          & Illustrative Content                   & \begin{tabular}[c]{@{}l@{}}Supportive information \\ provided to illustrate \\ the solution\end{tabular}       & \begin{tabular}[c]{@{}l@{}}Code, Outcome, \\ Expected Outcome\end{tabular}                &                           \\ \hline
\end{tabular}
 \caption{Structural aspects extracted from \so questions and associated accepted answers. For a given aspect, we present its definition and possible associated values.}
 \label{tab:aspects-so}
\end{table}
\end{landscape}

\subsubsection{\revOne{RQ3: Analyzing General and Domain-Specific Impact of Posting Activity on \so}}
To assess the overall impact of \cgpt's release on question, answer, and comment frequency, we grouped the frequency of each element by the day they were posted and then performed our analysis. 
\req{To evaluate the impact on domains, we adopt a similar approach by grouping the questions based on the challenging domains reported in RQ2 (\textit{programming languages}, and \textit{frameworks and libraries}). 
For each domain, we manually searched for the top 10 most cited tags, which were the same before and after \cgpt's release.
Then, for each tag, we compute the frequency at which they were used daily.
Although \textit{Frameworks and Libraries} are classified as one single domain, we explore them individually, selecting the ten most cited frameworks and libraries.} 

Knowing that our data did not follow a normal distribution, as verified by the Shapiro-Wilk test \citep{shapiro1965analysis},\footnote{\textit{p-value} \textless{} 0.01} we employed the non-parametric Wilcoxon test to compare the distributions of each evaluated group, \revOne{using a significance level of $\alpha = 0.05$}~\citep{10.1214/aoms/1177730491}.

\subsubsection{\revOne{RQ4: Checking Impact on User's Activity Patterns in \so}}
\rev{Based on the number of users active before and after the release of \cgpt, we first investigate the impact on the different types of users (questioners, respondents, and commentators).}
\rev{Different from the previous RQ, in this one, we focused only on the data before and after release.}
\revOne{Knowing that our primary goal is to evaluate the impact of ChatGPT's release on user activity, to ensure a focused comparison, we analyzed data from five months before and up to one year after its release (after and one-year milestone). 
While we acknowledge that external factors could influence user activity both before and after the release, restricting the pre-ChatGPT period helps minimize historical biases unrelated to ChatGPT, such as long-term platform trends, policy changes, technologies released, or shifts in users' activity patterns. 
While post-release activity may also be affected by other factors, by adopting a one-year window, we believe we can ensure that we analyze both immediate and sustained changes, making it more likely that observed trends are linked to ChatGPT’s introduction rather than unrelated fluctuations.}

Moving on, we first compute the IDs of the users associated with the questions and then compare which users were active or inactive after \cgpt's release. 
Second, we further compare the active users after release based on the creation date of their accounts. 
Third, based on the different types of users we observe (reminiscent and new users), we explore how users appeal for support on \so, regarding questions previously asked to \cgpt. 
This way, we mine all questions that refer to \cgpt and its varying terminologies on the question's title and body (\texttt{chat-gpt}, \texttt{chat gpt}, and \texttt{gpt}).\footnote{We did the same for \llamatwo, but we did not observe questions targeting it.}

\req{Initially, by analyzing all the \num{578115} questions after ChatGPT release (see Table \ref{tab:mining}), which also include questions without answers, we observe that 2193 questions mention ChatGPT in their contents (title or body). Next, checking the creation date of the questioners’ accounts, we observe that \num{1439} questions were asked by
reminiscent users, while 731 questions were asked by new ones. 
The remaining questions were asked anonymously or by users no longer with active accounts in \so, so we could not check whether the questions were associated with reminiscent or new users. 
For each group, we randomly select a subsample of questions to be manually analyzed (91 and 86 questions, respectively), aiming for a 95\% confidence level and a 10\% margin of error. 
Then, we manually analyze each question, extracting the information reported in Table \ref{tab:aspects-so}.
For the current analysis, we focus on the question and the associated accepted answer, as we are also interested in how the respondents would behave when facing a problem that LLMs were previously asked.
}

\section{Empirical Findings}
\label{sec:results}
This section presents the results of our empirical study.

\subsection{How does the reliability of answers generated by \cgpt and \llamatwo compare to those provided by
Stack Overflow users?}
\label{sec:rq2}

\req{In our first research question, after selecting \so questions and prompting LLMs to address them, we aim to check their reliability based on textual (cosine metric) and semantic similarity.}

\begin{figure}
    \centering
    \begin{subfigure}{0.49\columnwidth}
        \centering
        \includegraphics[width=\linewidth]{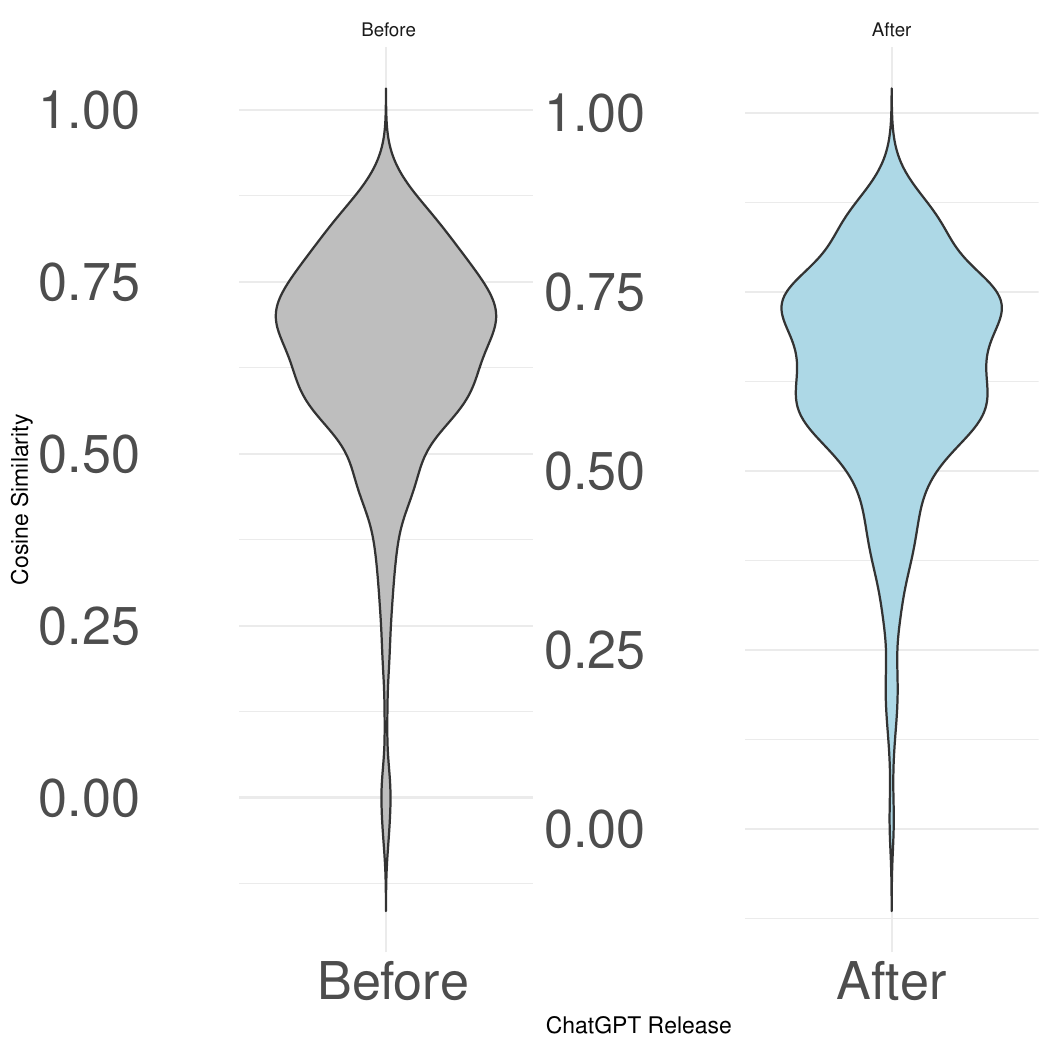}
        \caption{\cgpt}
        \label{fig:similarity-gpt}
    \end{subfigure}
    \begin{subfigure}{0.49\columnwidth}
        \centering
        \includegraphics[width=\linewidth]{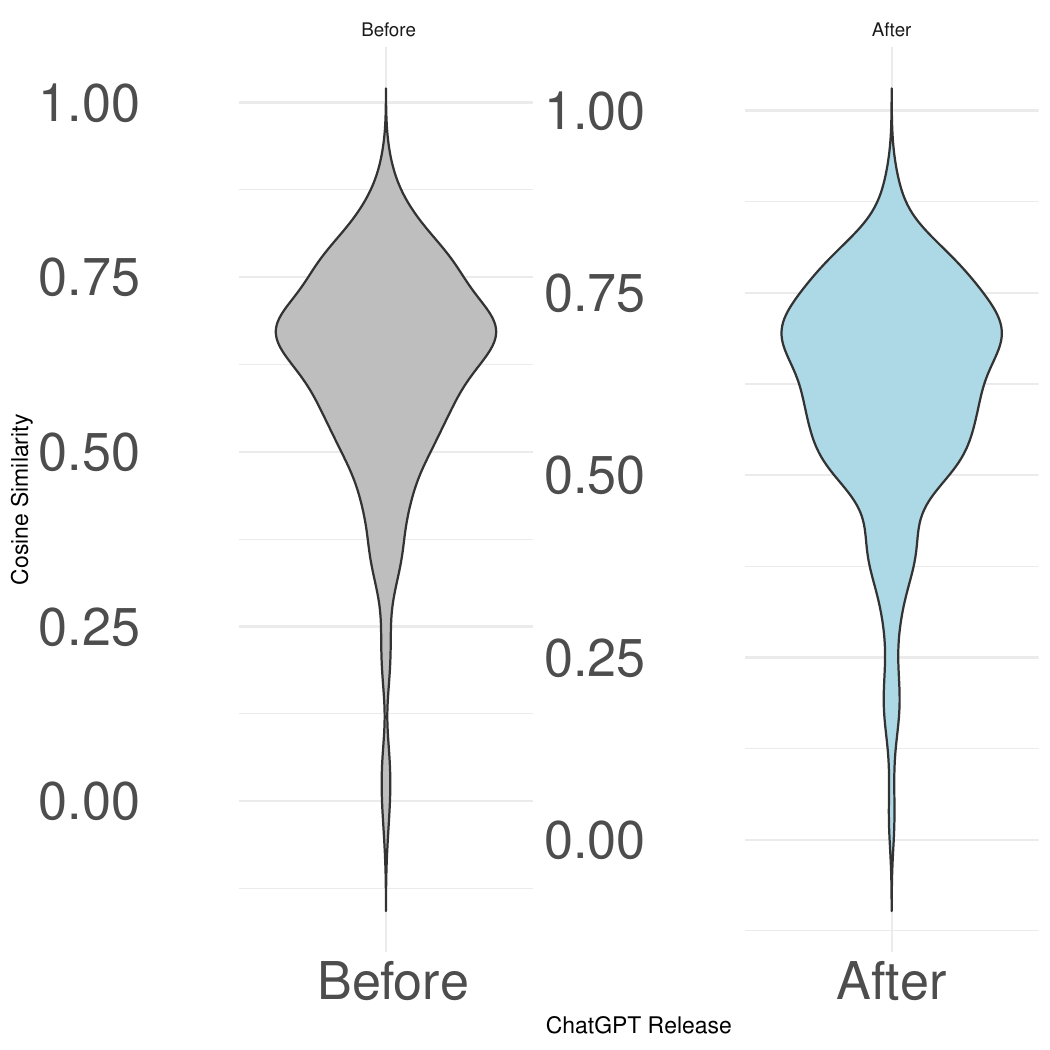}
        \caption{\llama}
        \label{fig:similarity-llama}
    \end{subfigure}
    \caption{Distribution of Cosine Similarity computed by each Large Language Model. The metric ranges from -1 to 1, representing dissimilarity and similarity, respectively.}
    \label{fig:similarity-violin-plot}
\end{figure}

\subsubsection*{Textual Similarity Analysis}

Regarding the textual similarity, Figure~\ref{fig:similarity-violin-plot} shows the distribution of the cosine metric across all prompted questions for both \cgpt and \llamatwo. 
Specifically for \cgpt, overall, both plots present similar densities (Figure \ref{fig:similarity-gpt}, i.e., before and after \cgpt release), indicating that there is no significant difference regarding the similarity between \so and \cgpt answers. 
By computing the mean for each time interval, we could validate our initial insight (0.644 and 0.640, pre- and post-release, respectively).
Further comparing them, we report no significant statistical difference between both distributions (\textit{Wilcoxon}, \textit{p-value} = 0.59).
\req{By reporting such overall similar answers, \cgpt shows its capability to address tech questions while adopting a similar behavior compared to humans at the textual level. 
Furthermore, we may highlight that such observation must consider that \cgpt was trained on publicly available data until September 2021, eliminating possible associated bias.\footnote{\url{https://platform.openai.com/docs/models/gpt-3.5-turbo}}}  

For \llama, the plots also present a similar density, observed by the average reported (Figure \ref{fig:similarity-llama}, 0.616 and 0.619, before and after \cgpt release, respectively), with a slightly more uniform similarity distribution for the ``after \cgpt release'' plot.
We also observed no significant statistical difference between both distributions (\textit{Wilcoxon} test, \textit{p-value} = 0.71), 
indicating that \llamatwo presented the same behavior for both sets of questions.
Different from \cgpt, this result might be unexpected considering that, though most of the data used to train \llamatwo were dated until September 2022, some tuning data was further required, dated up to July 2023, which might introduce bias in our results due to memorization~\citep{carlini2022quantifying}\footnote{\url{https://github.com/facebookresearch/llama/blob/main/MODEL_CARD.md\#training-data}}.

Since no significant difference was observed for time interval metrics based on each LLM, we decided to compare the LLMs by aggregating their data. 
As a result, \llamatwo presented an overall lower mean when compared to \cgpt (0.617 and 0.642, respectively).
A \textit{Wilcoxon} test reports that \cgpt was statistically superior to \llamatwo (\textit{p-value} \textless 0.01) but with a small effect size (Cliff's delta, 0.1).
Hence, we conclude that though \cgpt was statistically superior, \llamatwo was close to achieving similar results, representing an alternative, free, and public option for the tech community.
\req{To delve into the capabilities of the LLMs evaluated, we discuss some examples as follows.}

\noindent
\textbf{Before \cgpt's release.}
Considering the question with the highest similarity by \cgpt ($\theta$ = 0.92), the user places a question about a programming task in Java, asking about \emph{How to get all keys whose values are null in Java 8 using Map}\footnote{\url{https://stackoverflow.com/questions/73687017/}}.
An initial code snippet is provided to illustrate the task, presenting a \texttt{HashMap} with five \texttt{keys} representing colors associated with their HTML codes (two of which are \emph{null}).
Figures ~\ref{code:first_question_so}, ~\ref{code:first_question_cgpt}, and ~\ref{code:first_question_llama} present the answers provided by a human on \so, and generated by \cgpt and \llamatwo, respectively. 

\begin{figure*}[h!]
    \begin{subfigure}{\textwidth}
        \centering
        \begin{lstlisting}[language=Java, numbers=left]
Map<String, String> h;
h = new HashMap<>();

List<String> collect = h.keySet()
  .stream()
  .filter(e -> Objects.isNull(h.get(e)))
  .collect(Collectors.toList());
    \end{lstlisting}
        \caption{Answer generated by a human on SO.}
        \label{code:first_question_so}
    \end{subfigure}
    \begin{subfigure}{\textwidth}
        \centering  
        \begin{lstlisting}[language=Java, numbers=left]
        Map <String, String> h;
h = new HashMap<>();

List<String> collect = h.entrySet()
  .stream()
  .filter(e -> e.getValue() == null).map(Map.Entry::getKey)
  .collect(Collectors.toList());
        \end{lstlisting}
        \caption{Answer generated by \cgpt.}
        \label{code:first_question_cgpt}
    \end{subfigure}
    \begin{subfigure}{\textwidth}
        \centering
         \begin{lstlisting}[language=Java, numbers=left]
         Map <String, String> h;
h = new HashMap<>();

List<String> collect = h.values()
  .stream()
  .filter(e -> e == null)
  .collect(Collectors.toList());
         \end{lstlisting}
        \caption{Answer generated by \llama.}
        \label{code:first_question_llama}
    \end{subfigure}
      \caption{Three different versions of answers to a \so question.}
    \label{code:first_question}
\end{figure*}

Both \so and \cgpt answers address the user's request but adopt slightly different approaches.
Initially, they adopt a \texttt{stream} object to process the \texttt{HashMap} elements; while the human answer works only with the set of \texttt{keys} (line 4), \cgpt opts to work with the entire \texttt{set} (line 4).
Next, to check whether the \texttt{value} associated with a \texttt{key} is \texttt{null}, the human answer first requests the value associated with a key to \texttt{h} and then checks whether the returned \texttt{value} is \texttt{null} by calling the method \texttt{Objects.isNull} (line 6).
\cgpt directly asks for the \texttt{value} associated with each set element, and for the applicable cases, it gets the associated \texttt{keys} (line 6).
Finally, the selected \texttt{keys} are grouped similarly in a \texttt{List} for both cases (line 7).

However, \llamatwo's solution does not work correctly ($\theta$ = 0.89).
Indeed, the solution returns the \texttt{values} associated with the \texttt{keys} instead of just the \texttt{keys} themselves.
It occurs because instead of initially getting the \texttt{HashMap} keys, the proposed solution gets the set of \texttt{values} (line 4).
Then, for each \texttt{value}, it is checked whether they are \texttt{null} (line 6).
For the valid cases, they are stored in the \texttt{list} used to save the results (line 7).
\req{Such a result shows that though \llamatwo was close to addressing the requested task, the reported answer requires further adjustments. 
An experienced user could run the code and detect the misbehavior. Next, with a few updates, the code could be fixed and behave as expected.}

\noindent
\textbf{After \cgpt's release.}
Now consider another question with the highest similarity achieved by \cgpt dated after its release ($\theta$ = 0.92).
In this example, the user asks
for assistance regarding responsive design.\footnote{\url{https://stackoverflow.com/questions/75494817/}}
For that, the user provides some \texttt{CSS} code, defining different screen and font sizes, and based on screen size, it expects the elements to be adjusted.
The \so respondent provides four possible solutions for the problem:\footnote{\url{https://stackoverflow.com/a/75494849/5141439}}

\begin{quoting}
    \emph{\dcircle{1} Make sure that the .my-element class is being applied to the correct element in your HTML. \dcircle{2} Check that there are no other styles elsewhere in your CSS that might be overriding the font size changes made by the media queries. \dcircle{3} Try adding the !important declaration to the font-size property in each media query. \dcircle{4} Verify that your browser window size is within the range specified.}
\end{quoting}

\cgpt's answer overlaps three:
\dcircle{1} double-checking whether the expected \texttt{HTML} element is correctly targeting by the specified property;
\dcircle{2} possible style conflicts overriding the expected font sizes; and 
\dcircle{3} checking whether the browser window size is valid.
Furthermore, the \so respondent advises the user to use the \texttt{!important} declaration, ensuring that property is prioritized over others (style conflicts).
\cgpt adopts a more factual solution, asking the user to check whether the \texttt{CSS} file is correctly linked to the expected \texttt{HTML} file.

\llamatwo takes a different route compared to \cgpt and the original answer, by exploring solutions involving exclusively the code given as input ($\theta$ = 0.72).
The solution proposed is related to possible properties being overridden by others,  which \cgpt and \so answers also briefly discussed.
To address the issue, \llamatwo suggests replacing the usage of the property \texttt{max-width} for \texttt{min-width}, or the previous two properties for just \texttt{break-point}.
\req{These examples show the ability of \cgpt and \llamatwo to address different subjects, with a high similarity when compared to human answers, leading us to observe a good level of reliability.}

\subsubsection*{Semantic Similarity Analysis}
Evaluating reliability relying exclusively on textual similarity might introduce biases in our results, as we briefly discuss when presenting the question addressed by \llamatwo, which was not properly correct.
The opposite scenario can also occur; for example, a user asks for instructions about exporting a fillable textbox PDF file using PowerBI. 
The accepted answer reports that the required action is not possible to be done\footnote{\url{https://stackoverflow.com/questions/74639660/}}.
\cgpt correctly informs that the required task is currently impossible to do. 
However, the LLM provides further alternatives that might support the user, resulting in a low textual similarity ($\theta$ = 0.10).
We may conclude that textual similarity represents a good metric for reporting high similarity between the answers but not for reporting low similarity, as false negatives might take place.
To address this threat, \revOne{we also relied on LLMs’ judgments to compute the semantic similarity between answers and analyzed the level of agreement in these judgments.}
Overall, \llamatwo reported a more positive evaluation with only 7.4\% of the answers with low similarity, whereas \cgpt reported a frequency of 22.1\%.
This result goes the opposite of our manual classification, where the frequency of low similarity was around 44\% of our sample.

When computing the Krippendorff alpha between \cgpt and \llamatwo, we observe a coefficient of 0.02.
\req{Although we have used the LLMs with their default configurations, we observe that their ratings are not in good agreement, leading us to conclude that their ratings may not be reliable for the given data}.
When computing the Krippendorff alpha for multiple runs using \cgpt, we observed more consistent results but still had a low coefficient (0.34).
This result shows that the LLMs, even on different runs of the same model, do not present consistent and stable results regarding judging the semantic similarity of the answers.
\req{We believe that the temperature and advanced approaches might play a role in our findings.
However, even in such conditions, we might expect some inconsistency when judging semantic similarity.}
Finally, we also observe the same low similarity when evaluating the agreement of the LLMs with our manual analysis (Krippendorff alpha 0.26 and -0.21, \llama and \cgpt, respectively).
As a result, \req{we can conclude that possible conclusions we draw here based on this semantic similarity would be biased by the random behavior of the LLMs (see Section~\ref{sec:threats})}.

\subsubsection*{Sentiment Analysis}
Regarding the sentiment analysis performed, we observe that most of the answers provided by \so users and LLMs present a neutral sentiment associated (96\%, 97\%, and 79\%, \so, \cgpt, and \llama, respectively).
Unlike other web forums, \so users are used to keeping a formal, neutral, and direct way of speaking, which could be misinterpreted as demanding (negative sentiment), especially when listing a sequence of steps to be followed. 
When comparing the answers, we observe that \cgpt reports a more neutral sentiment compared to the original answers from \so users (\textit{Wilcoxon} test, \textit{p-value} \textless 0.01).
Furthermore, \cgpt did not report any answer with a negative sentiment, while 13 user answers were classified with negative sentiment.
Such a result reinforces the ability of \cgpt to adapt itself when answering questions, specifically in a tech context evaluated here.

Comparing \llama answers with those generated by \cgpt and \so users, we observe a significant statistical difference, revealing that \llamatwo report a less neutral sentiment (\textit{Wilcoxon} test, \textit{p-value} \textless 0.01 and 0.01, respectively).
However, unlike the previous answers, \llamatwo reports more answers with positive sentiment (20\%).
These results show that, in its current version, \llamatwo consistently adopts a positive-neutral sentiment when prompted by users. 

\highlighttwo{
\textbf{RQ1 answer:}
Checking the reliability based on textual similarity is statistically consistent. 
Both before and after \cgpt's release, the generated answers show a high degree of similarity to the accepted answers on \so, as evidenced by the textual similarity (cosine metric).
Although textual similarity seems to be a good metric for reporting answers with high similarity, this metric is not applicable for reporting low similarity, as false negatives might occur.
Finally, the adopted semantic similarity metric based on LLMs classification also does not present an applicable metric for checking similarity among the answers.
Although \cgpt presents a more neutral sentiment when generating their answers, \llamatwo presents
a more positive sentiment.
}

\subsection{RQ2: What specific domains pose challenges for \cgpt and \llamatwo when generating reliable and accurate answers?}
\label{sec:rq3}

\begin{figure}
    \centering
    \includegraphics[width=1\linewidth]{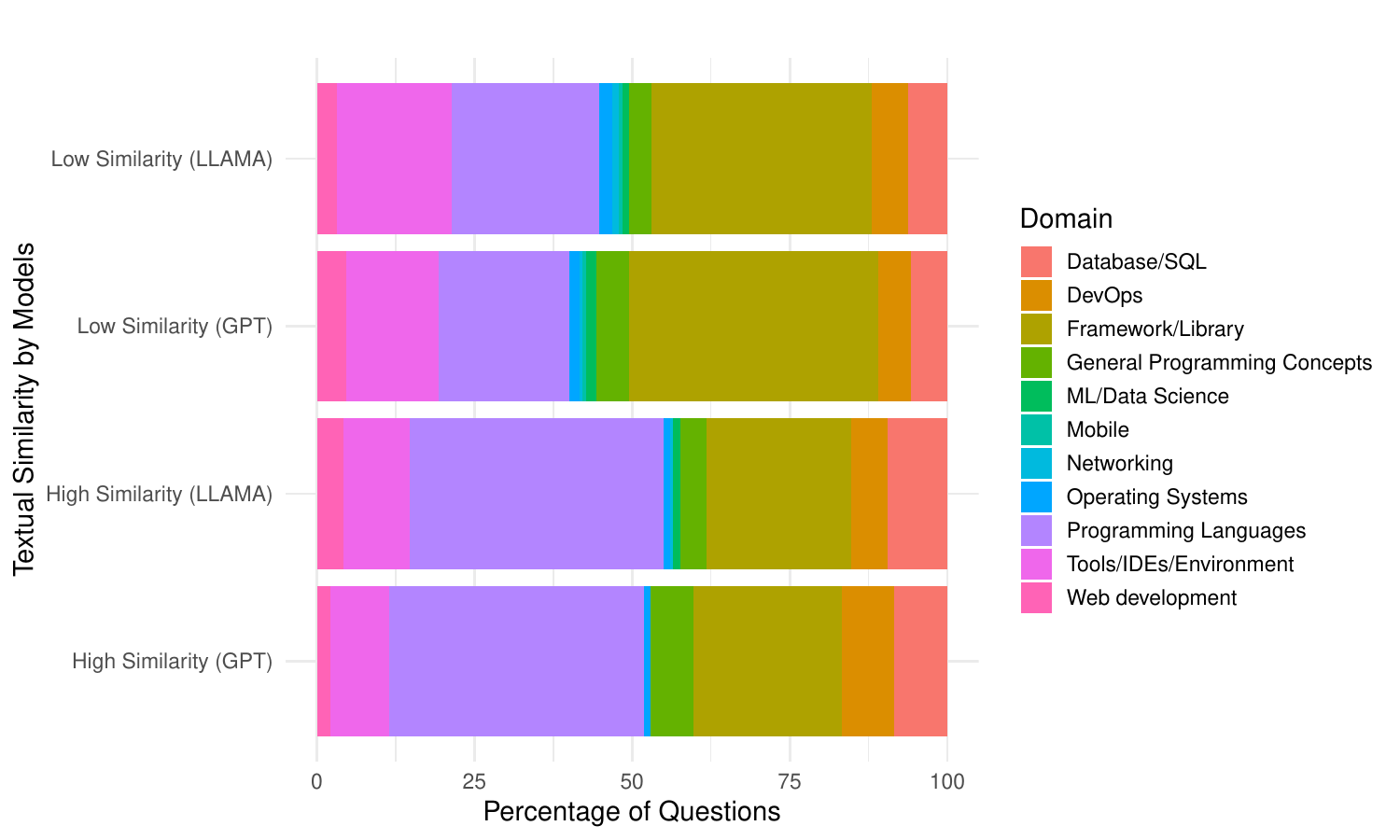}
    \caption{\revOne{Distribution of question domains for high- and low-similarity answers generated by LLMs (\cgpt and \llamatwo)}.}
    \label{fig:domain_distribution}
\end{figure}

This research question focuses on understanding the associated challenges faced when generating these answers and the factors that might lead the LLMs to generate more accurate answers.
Next, we discuss each aspect individually. 

\subsubsection*{Challenges for LLMs}
To report the challenges faced by LLMs, we manually analyzed the questions associated with high and low textual similarity (see Section \ref{sec:manual-low-high}).
Regarding the domain of the questions classified with low textual similarity for LLMs (Figure \ref{fig:domain_distribution}), we reach the same conclusions for both LLMs. 
\textit{Frameworks and Libraries} represent the most challenging domain for the LLMs evaluated here (39.6\% and 34.9\%, \cgpt and \llama, respectively), followed by \textit{Programming Languages} (20.8\% and 23.4\%).
When looking at the different groups of questions with the low textual similarity associated with each LLM, we observe that they share 64\% of the same questions, revealing that despite the structural differences between the models, they partially share the same limitations when addressing these domains.
Furthermore, less than 20\% of all generated answers by both models presented a similarity of less than 0.5 (14\% and 16\%, for \cgpt and \llama, respectively).
This small frequency shows that even when the LLMs fail to create a more similar answer, they do not deviate from the original direction of the accepted answers.

On the other hand, for the questions with high textual similarity, the most promising domains are Programming Languages (40.5\% and 40.3\%), followed by \textit{Frameworks and Libraries} (23.7\% and 23\%). 
Although the \textit{Frameworks and Libraries} domain is previously reported as a challenging domain for LLMs, we must discuss how the type of questions associated with each group plays an important role here. 
When users ask for specific, popular, and standard features of these dependencies, the LLMs achieve a high similarity; however, when the questions address \revOne{general conventions, standard practices or common knowledge adopted by the community which may not be explicitly documented,} or unpopular features about using these external dependencies, a low similarity is observed.
For example, a user asks for support when failing to authenticate to an application using valid credentials. 
The solution accepted guides the user to hash the password when creating a new user in the application:\footnote{\url{https://stackoverflow.com/questions/75727923/}}

\begin{quoting}
    \emph{You need to hash the password before creating the user […] You will need to create the user again since you will create a user with a \textbf{hashed} password to let the authentication succeed.}
\end{quoting}

In its generated answer, \cgpt even warns about the user's credential and object creation steps without mentioning the actual cause.

\req{Further exploring the questions, we could observe some structural aspects that might bring some light to our observations. For example, when we look for external references linked to questions regarding framework and dependencies, we observe that the LLMs performed poorly for questions with more external references.
While \cgpt and \llama had to deal with 10 and 13 questions with external references when reporting low similarity, they had to deal with only 2 and 4 questions, respectively, when reporting higher similarity.}
\revOne{For example, consider a user facing an issue when logging out in a VueJS app using Supertokens for authentication. 
Even after clicking the Logout button, the session data remained (\texttt{userInfo}), and the user still had an active session. 
Attached to the question, the user also reports a link holding a code snippet provided by SuperTokens.\footnote{
Currently, the provided link is unavailable; however, for each code snippet reported in the documentation, SuperTokens reports an associated explanation.}
When answering the question, the respondent clarifies that the browser was not saving session cookies because the frontend was running on localhost while the API was on 127.0.0.1, and cross-domain requests require HTTPS for browsers to store cookies properly.
Without these extra external details, the LLMs take a different route, mainly guiding the user to check the setup of the environment and sign-out options.}

\revOne{Although the advanced version of \cgpt can access external links using the web tool (\texttt{GPT-4-turbo}), allowing the model to perform searches and open URLs when needed, the evaluated models in this study do not have such a capability.}
\revOne{By linking external content, the questions  provide a less self-
contained context, and consequently, users are expected to read and follow these links for complete understanding. 
As a result, these questions are more challenging for LLMs to address, as the models must rely solely on the information explicitly provided in the prompt without access to any external supplementary content.}

\req{Another structural aspect observed is related to the level of detail provided in the questions. 
Although both groups provide code as supportive material, we observe that LLMs report higher similarity for questions supported not only by code but also by more supplementary data, like output, expected outcome, and error messages.
In this context, for questions with low similarity, we observe \cgpt and \llama dealing with only code for 79\% and 74\% of the questions, respectively, while for questions with high similarity, the rate was around 60\% and 62\%, respectively (\revOne{information collected as previously reported in Table \ref{tab:aspects-so}}).
These additional data provide more details about the problem, allowing the LLMs to have a better understanding of the target context.}

\req{We observe the same behavior regarding questions addressing programming language tasks when being supported exclusively by code. 
However, the difference was smaller than the previously discussed domain (for low similarity 78\% and 85\%, and for high similarity 72\% and 76\%, for \cgpt and \llama).
We believe that programming language questions are more direct as they involve only the aspects of the language itself.
For frameworks and libraries, further aspects might be considered, like environment setup and versioning.
However, further studies are required to better understand how these aspects could be better handled when prompting LLMs for related questions.}

\subsubsection*{Textual Similarity Correlation}

From all performed analyses, the results show only a statistically significant correlation between the number of tags adopted for a question and the similarity for the answers generated post \cgpt release (\textit{p-value} \textless 0.05).
The correlation coefficient shows a negative relationship with $\rho = -0.11$, meaning that while one variable increases, the other might decrease.
\req{This result indicates that when \cgpt is prompted to address problems related to one specific tag, it reports more reliable answers (high similarity). 
\revOne{While one might expect that providing more information (i.e., multiple tags) would improve reliability, we believe that it can have the opposite effect.}
For example, the two questions discussed in Section~\ref{sec:rq2} only adopt one single tag (\texttt{java-8} and \texttt{media-queries}, respectively).
\revOne{We believe that by restricting the question to one tag, the LLM can focus its reasoning on a well-defined topic, reducing the risk of generating off-topic responses.}
However, this negative relationship reveals that \cgpt might struggle when different contexts are present in the same question, as it has to arrange different topics to generate an answer.}
\revOne{When a question includes multiple tags, especially with unrelated concepts, the LLM may struggle to establish a clear reasoning path, potentially affecting the reliability of its answers.}
For instance, when a user asks for support regarding an unresolved reference, \revOne{they use} three tags, which are all related to each other (\texttt{android}, \texttt{android-livedata}, and \texttt{android-lifecycle})\footnote{\url{https://stackoverflow.com/questions/75449324/}}. 
For this question, \cgpt generated an answer opposite from the one provided by the \so user, achieving a low cosine similarity ($\theta$ = 0.37).

\highlighttwo{
\textbf{RQ2 answer:}
Although there are structural differences between \cgpt and \llamatwo, the LLMs partially share the same challenges. 
Questions about general conventions or unpopular features of \textit{Framework and Libraries} pose a challenge for the LLMs evaluated here, while \textit{Programming Language} questions are expected to be addressed appropriately. 
\req{Additional material linked to the questions might play a role in the similarity reached by the LLMs, like code snippets with additional related information.}
The number of tags associated with a question negatively correlates with textual similarity provided by \cgpt.
}

\subsection{RQ3: \revOne{What is the impact of \cgpt’s public release on \so posting activity?}}
\label{sec:rq1}

\begin{figure*}
    \begin{subfigure}{0.5\textwidth}
        \centering
        \includegraphics[width=\linewidth]{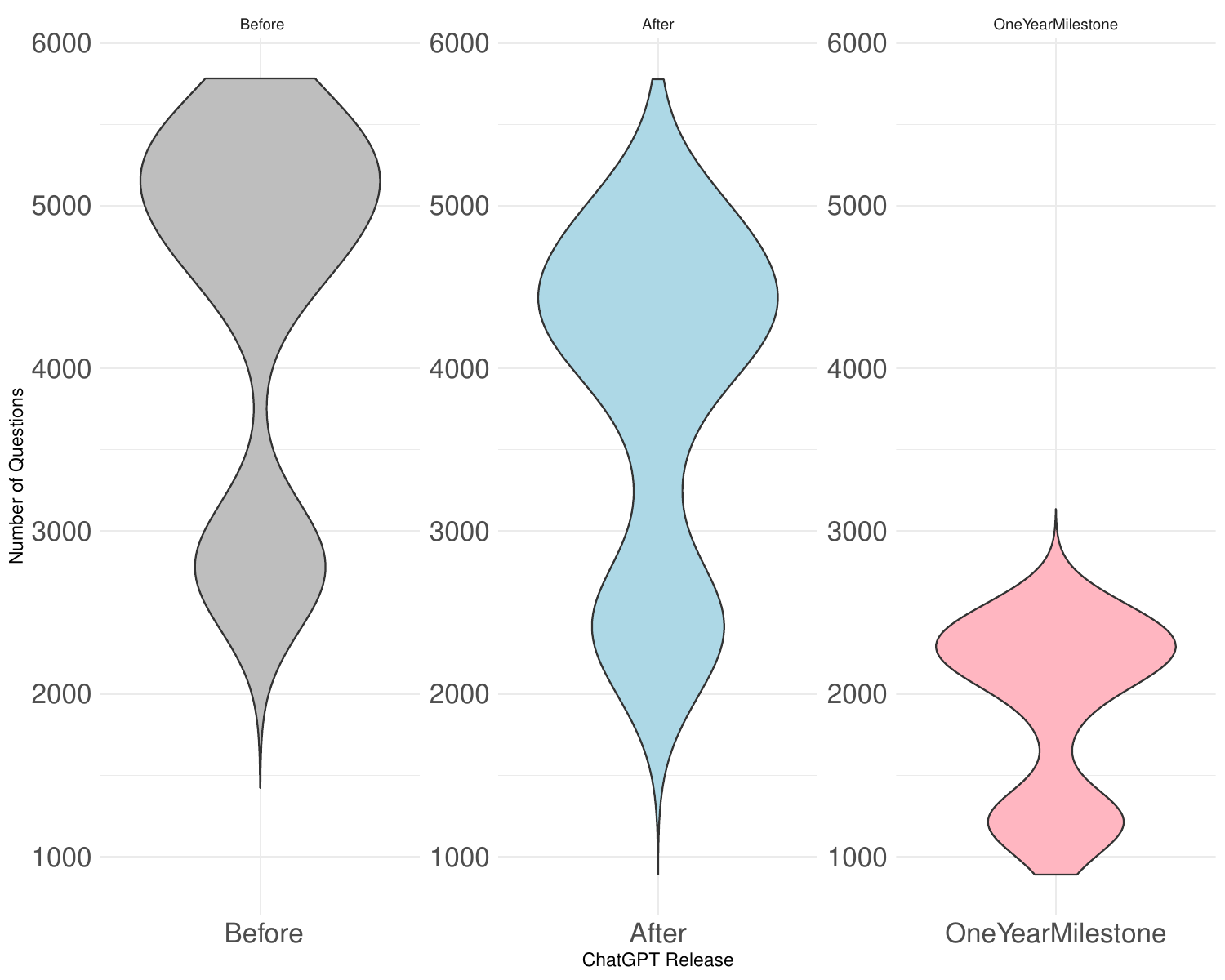}
        \caption{Questions}
        \label{fig:subfig1}
    \end{subfigure}
    \begin{subfigure}{0.5\textwidth}
        \centering
        \includegraphics[width=\linewidth]{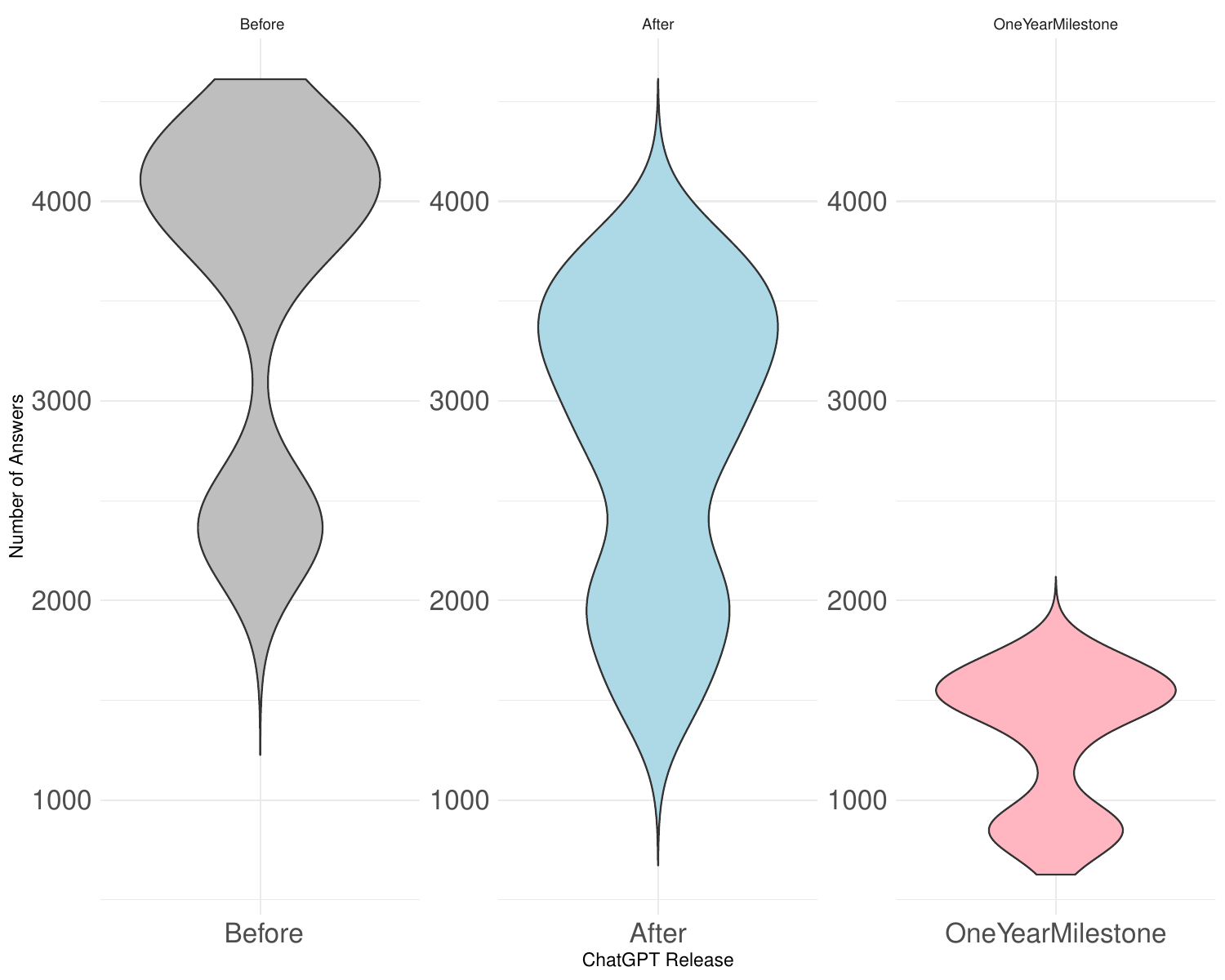}
        \caption{Answers}
        \label{fig:subfig2}
    \end{subfigure}
    \begin{subfigure}{0.5\textwidth}
        \centering
        \includegraphics[width=\linewidth]{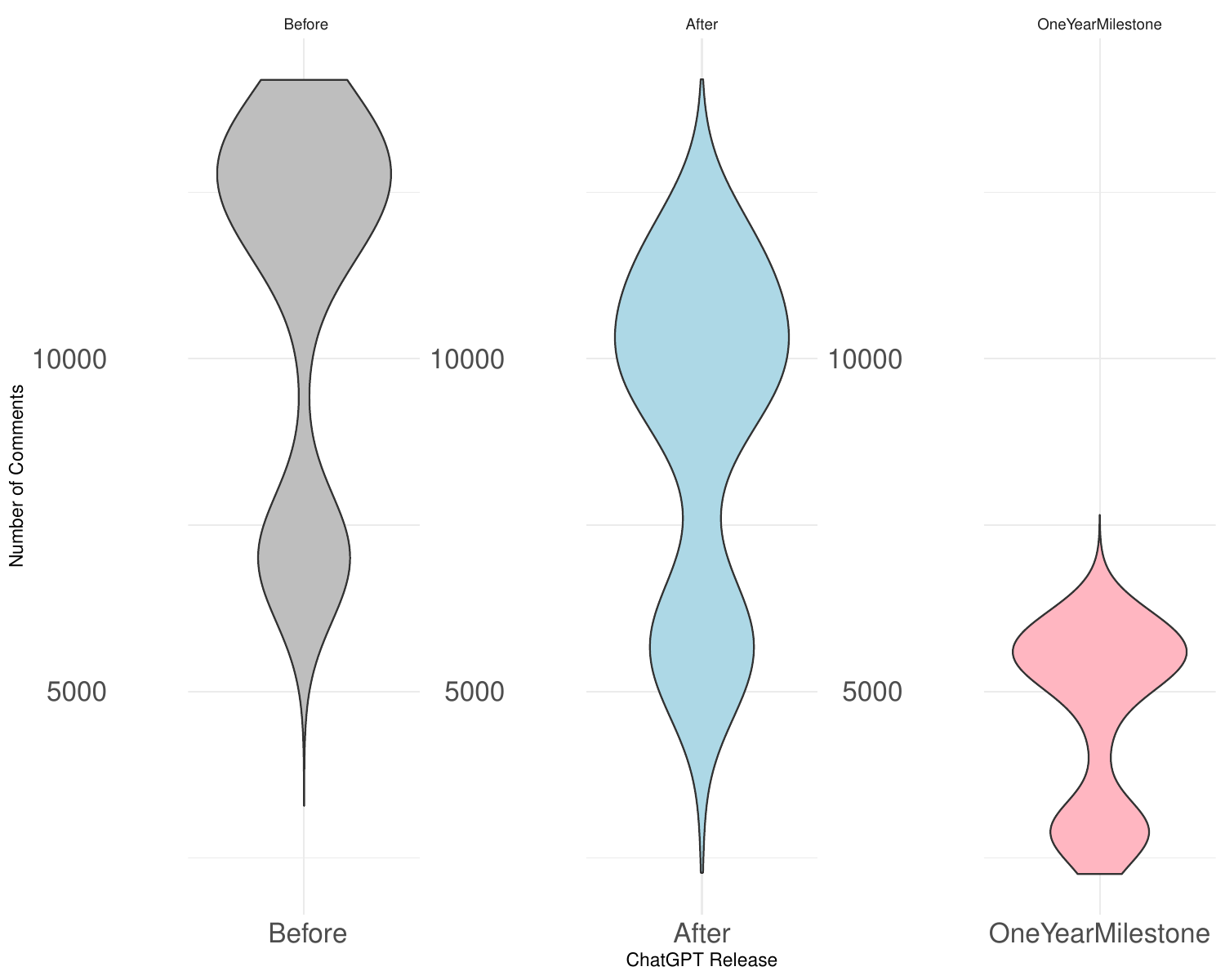}
        \caption{Comments}
        \label{fig:subfig3}
    \end{subfigure}
    \caption{\revOne{Violin plots comparing the distribution of the data across three time intervals: Before (July to November 2022), After (November 2023 to May 2023), and One-Year Milestone (November 2023 to May 2024).}}
    \label{fig:comparison-before-after}    
\end{figure*}

\begin{figure*}[!h]
    \begin{subfigure}{0.5\textwidth}
        \centering
        \includegraphics[width=\linewidth]{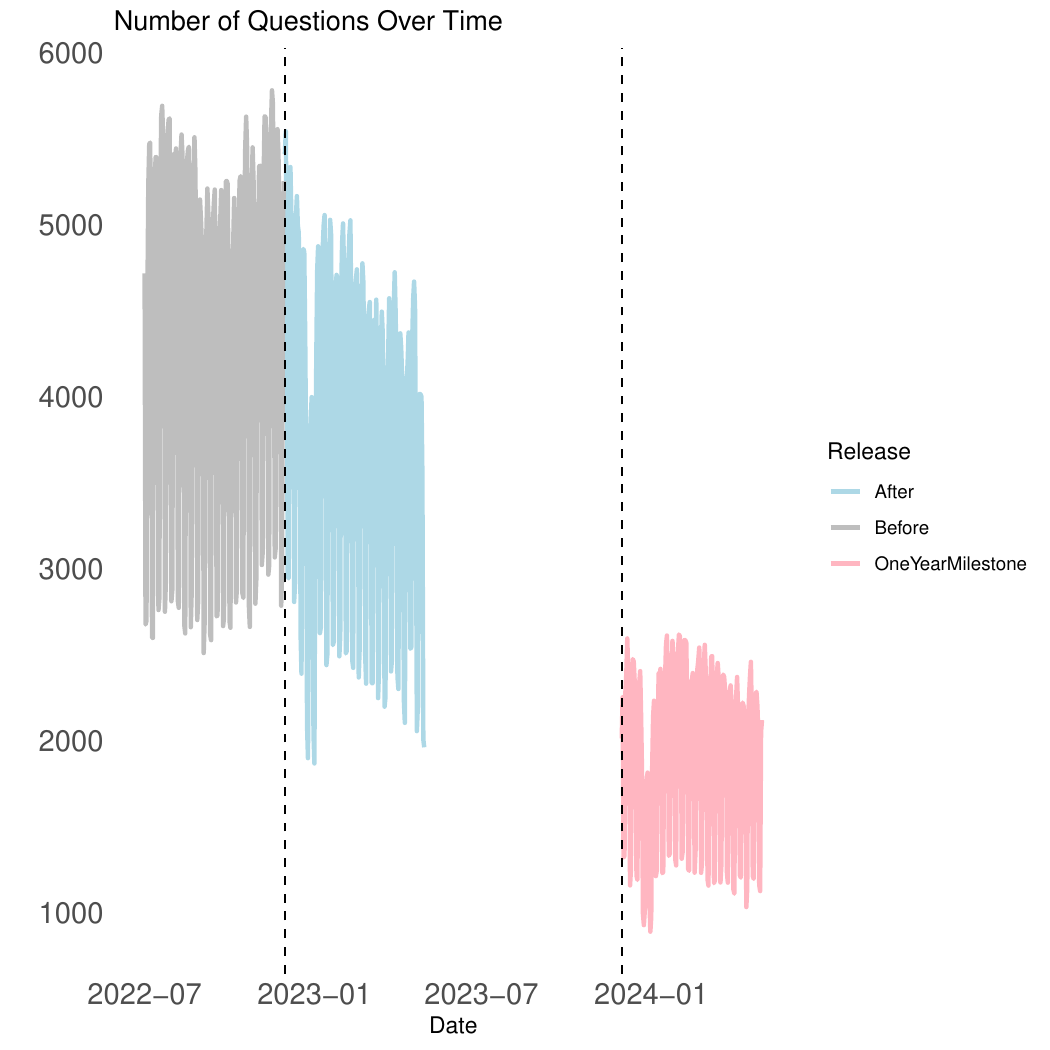}
        \caption{Questions}
        \label{fig:subfig2-1}
    \end{subfigure}
    \begin{subfigure}{0.5\textwidth}
        \centering
        \includegraphics[width=\linewidth]{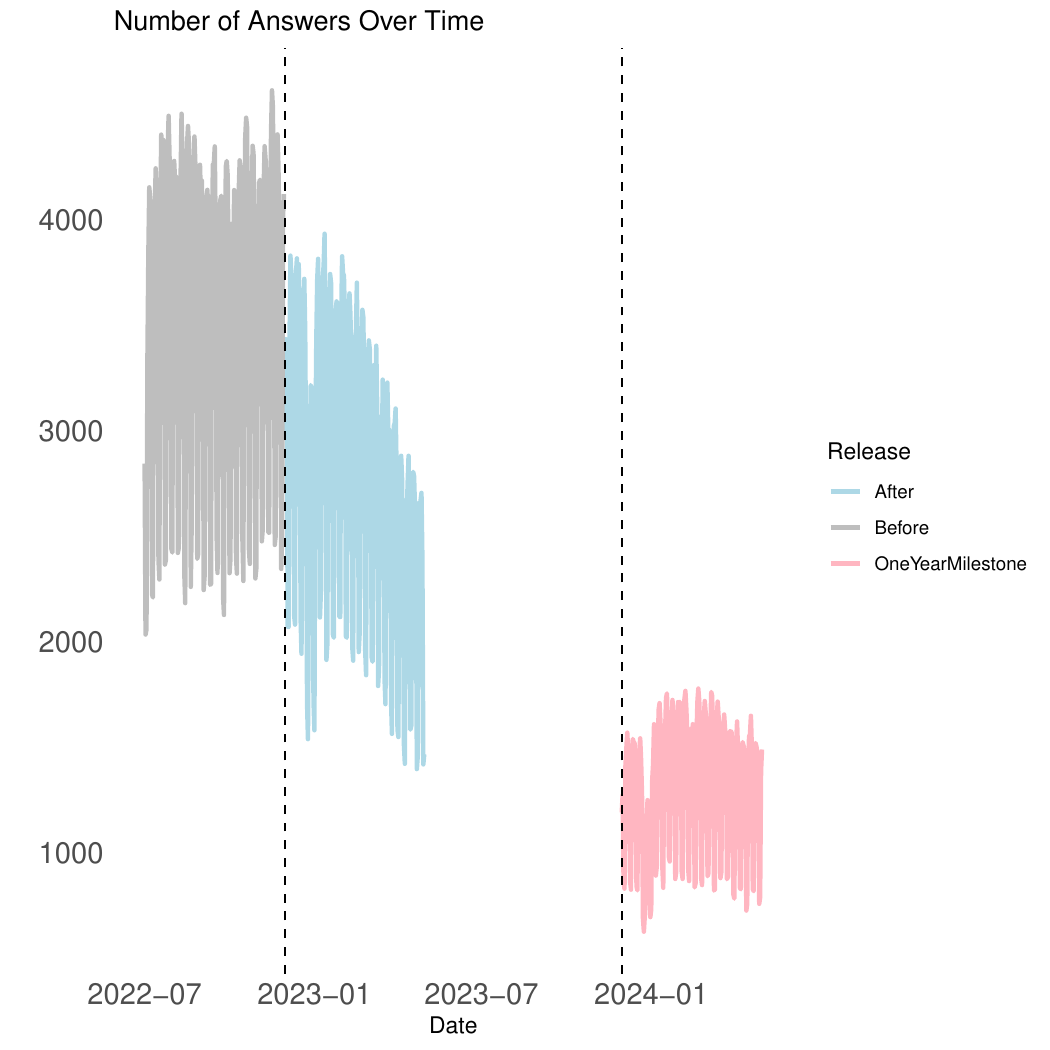}
        \caption{Answers}
        \label{fig:subfig2-2}
    \end{subfigure}
    \begin{subfigure}{0.5\textwidth}
        \centering
        \includegraphics[width=\linewidth]{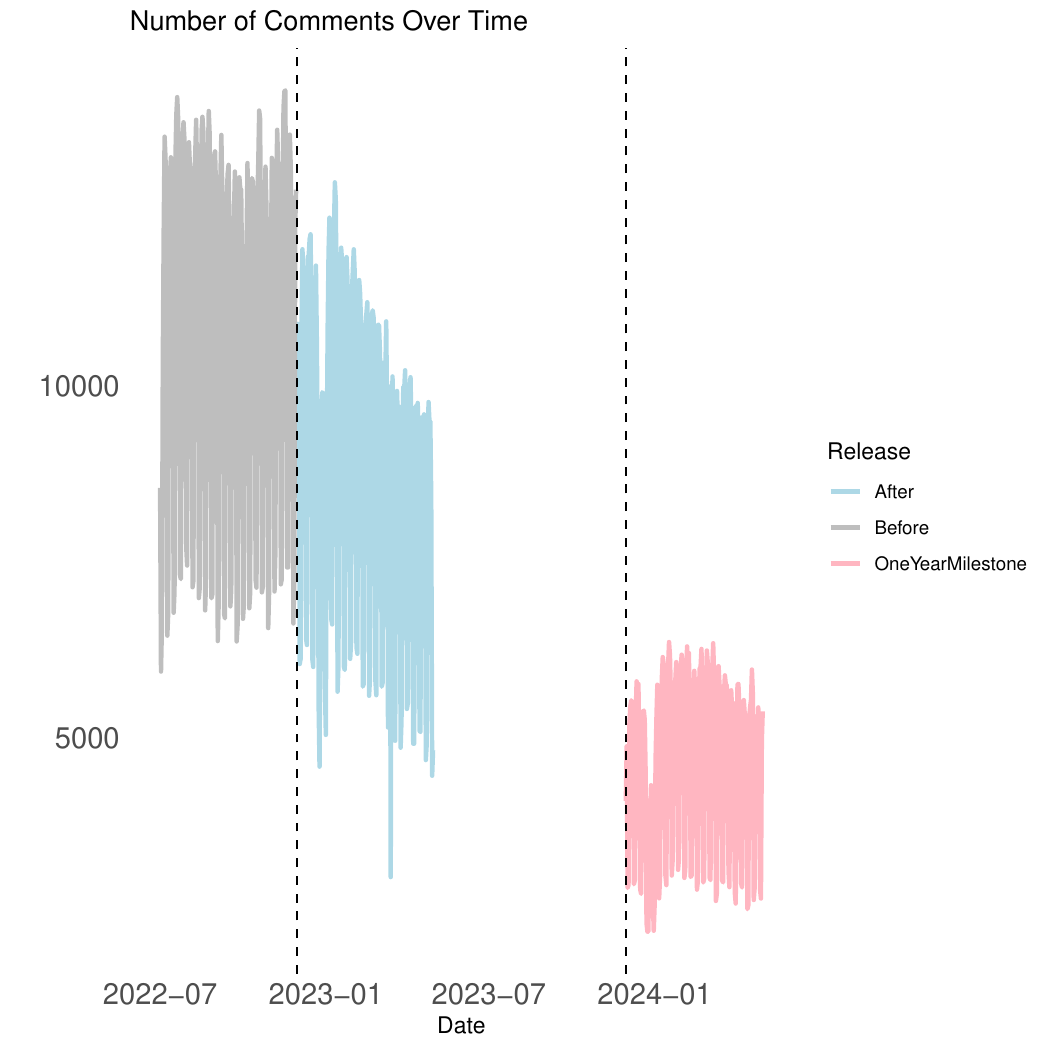}
        \caption{Comments}
        \label{fig:subfig2-3}
    \end{subfigure}
    \caption{
    \revOne{Distribution of posted content on \so (questions, answers, and comments). The gap between the \textit{After} and \textit{One-Year Milestone} intervals represents the period from May to November 2023, during which no data was collected.}}
    \label{fig:distribution-posted-contents}    
\end{figure*}

Our third research question investigates the impact of \cgpt's release on the content posted to \so.
First, we discuss the overall impact of posting and related activities, and then we discuss the impact on specific domains.

\subsubsection*{Impact on Posting Activities in \so}
\revOne{To address this research question, we first collected the relevant data, further comparing their frequency regarding the number of posted questions, answers, and comments before and after the introduction of \cgpt (see Table~\ref{tab:mining}).}
Notably, the number of submitted answers exhibits the most significant decline: 
\num{543533} answers before and \num{425391} answers after \cgpt (i.e., -22\%).
\req{After the one-year milestone, the decline is even more significant, with a drop of 52\% (from \num{425391} to \num{202326} answers).}
This decline has a subsequent impact on the number of questions with accepted answers, which dropped from \num{212775} before \cgpt to \num{157187} after \cgpt, and consequently to \num{78 262} after one year of the release (i.e., -26\% and -50\%, respectively).
Comments and questions follow closely, showing a slightly more significant decline after one year of the release (46\% and 48\%, respectively).

\revOne{To gain initial insights into potential trends regarding the decline, Figures~\ref{fig:comparison-before-after} and~\ref{fig:distribution-posted-contents} illustrate the content and temporal evolution of \so.} 
Figure~\ref{fig:comparison-before-after} reveals that before \cgpt's release, there was a higher density of posted content in the upper region of the violin plots. 
Just after the release, this concentration appears more evenly distributed across the plots, including the middle and lower regions. 
\req{Finally, after the one-year milestone, we observe a drastic decrease in the concentration of contents.}
Overall, the median daily count of submitted content is lower after \cgpt's introduction. 
Figure~\ref{fig:distribution-posted-contents} shows the decline in posted content over time.
Importantly, oscillating patterns present in the data before \cgpt's release continue to be observed (they correspond to weekends).

\revOne{Finally, to statistically assess whether there is a significant difference between the number of posted questions, answers, and comments before, just after, and the one-year milestone of \cgpt's release, we ran the \emph{Wilcoxon} rank sum test.} 
Our results show we can reject the null hypothesis for all the evaluated contents (\textit{p-value} \textless 0.01).
\req{Regarding the drop of contents after \cgpt's release, \revOne{we observe a moderate effect across all evaluated scenarios,\footnote{$r = 0.396, 0.491, 0.463$, for posting questions, answers, and comments, respectively.}} reinforcing the findings of previous related studies \citep{del2023large,burtch2023consequences}.
Regarding the decline of posted content after the one-year milestone, \revOne{we observe a large magnitudinal effect, showing a substantial impact on the posting activities,\footnote{$r = 0.745 , 0.801, 0.698$, for posting questions, answers, and comments, respectively.}} confirming our initial observations reported in Figures~\ref{fig:comparison-before-after} and \ref{fig:distribution-posted-contents}, and bringing evidence about the constant impact in \so.
While this reflects a significant decrease in posted content, innovations like OverflowAI may influence the shared data volume, as users are driven to consult the tool before publishing new content. 
This way, our findings cannot ensure a decline in \so views, as users might still be accessing \so but exploring previously posted content.
Further details are discussed in Section \ref{sec:discussion}.
}

\highlighttwo{
\textbf{RQ3 answer:}
After the introduction of \cgpt, we observe a statistically significant decline in users' activity on \so. 
This is evidenced by the reduced frequency of posted questions, answers, and comments usage, with the most pronounced decline observed after the one-year milestone following \cgpt's release. 
Hence, \cgpt's introduction has impacted the platform's dynamics, likely due to its ability to address many of the queries previously posted on \so.
}

\subsubsection{RQ3.1: How are different domains from \so impacted by the release of \cgpt?}
\label{sec:impact-domain}

\begin{figure}
    \centering
    \begin{subfigure}{\textwidth}
        \centering
        \includegraphics[width=0.6\textwidth]{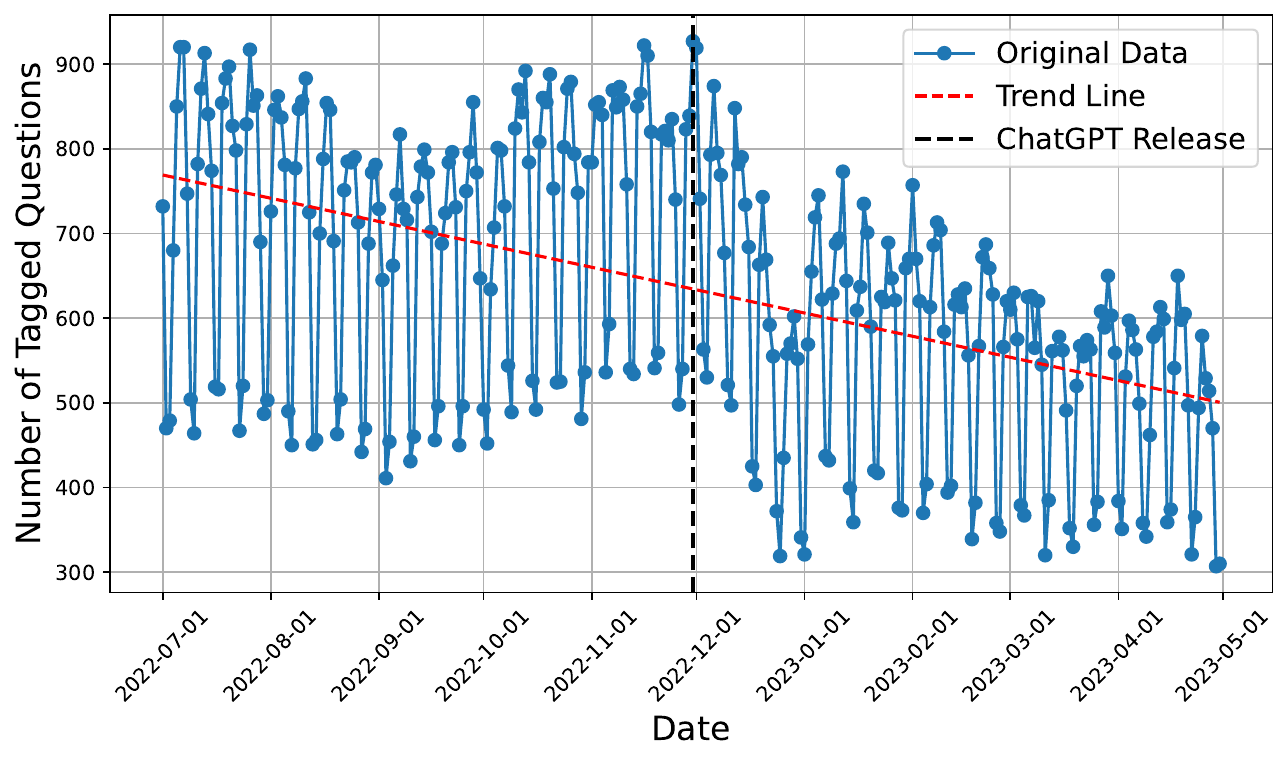}
        \caption{Python (Programming Language)}
        \label{fig:trend-python}
    \end{subfigure}
    
    \vspace{0.5cm} 
    
    \begin{subfigure}{\textwidth}
        \centering
        \includegraphics[width=0.6\textwidth]{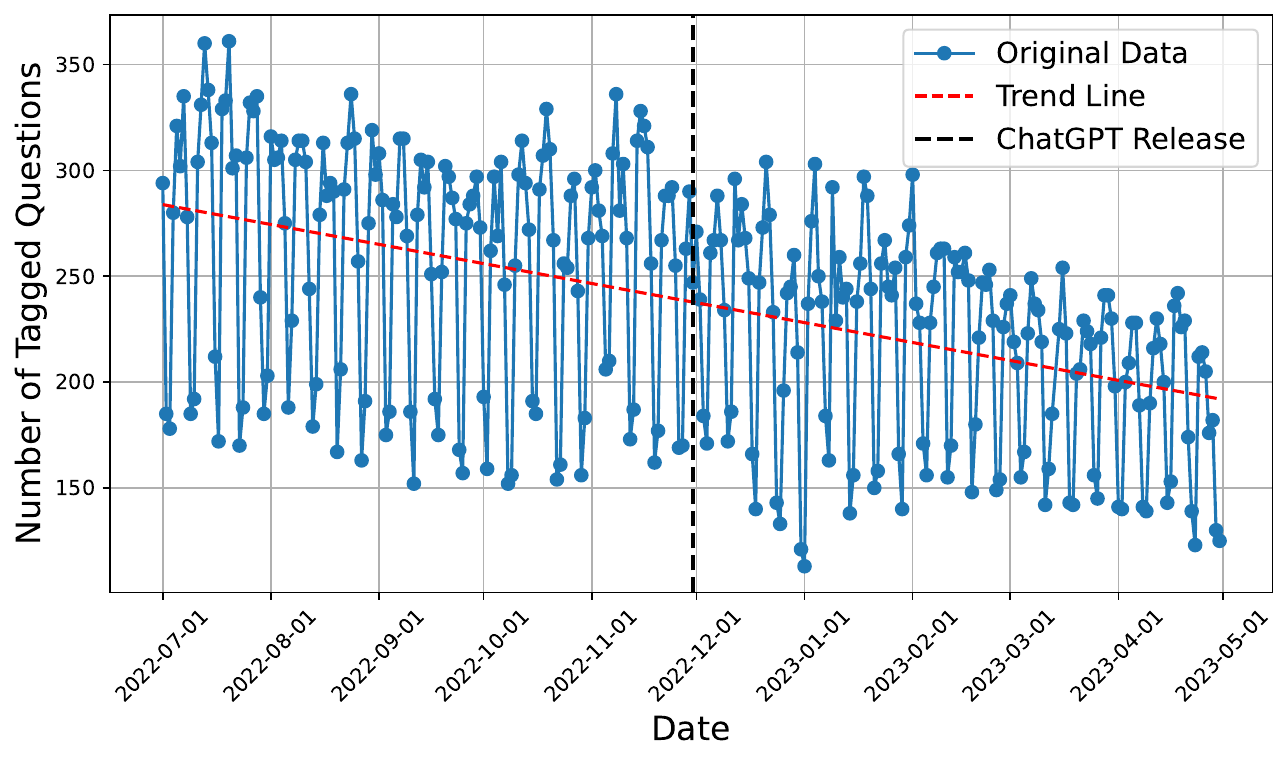}
        \caption{ReactJS (Framework)}
        \label{fig:trend-reactjs}
    \end{subfigure}
    
    \vspace{0.5cm} 
    
    \begin{subfigure}{\textwidth}
        \centering
        \includegraphics[width=0.6\textwidth]{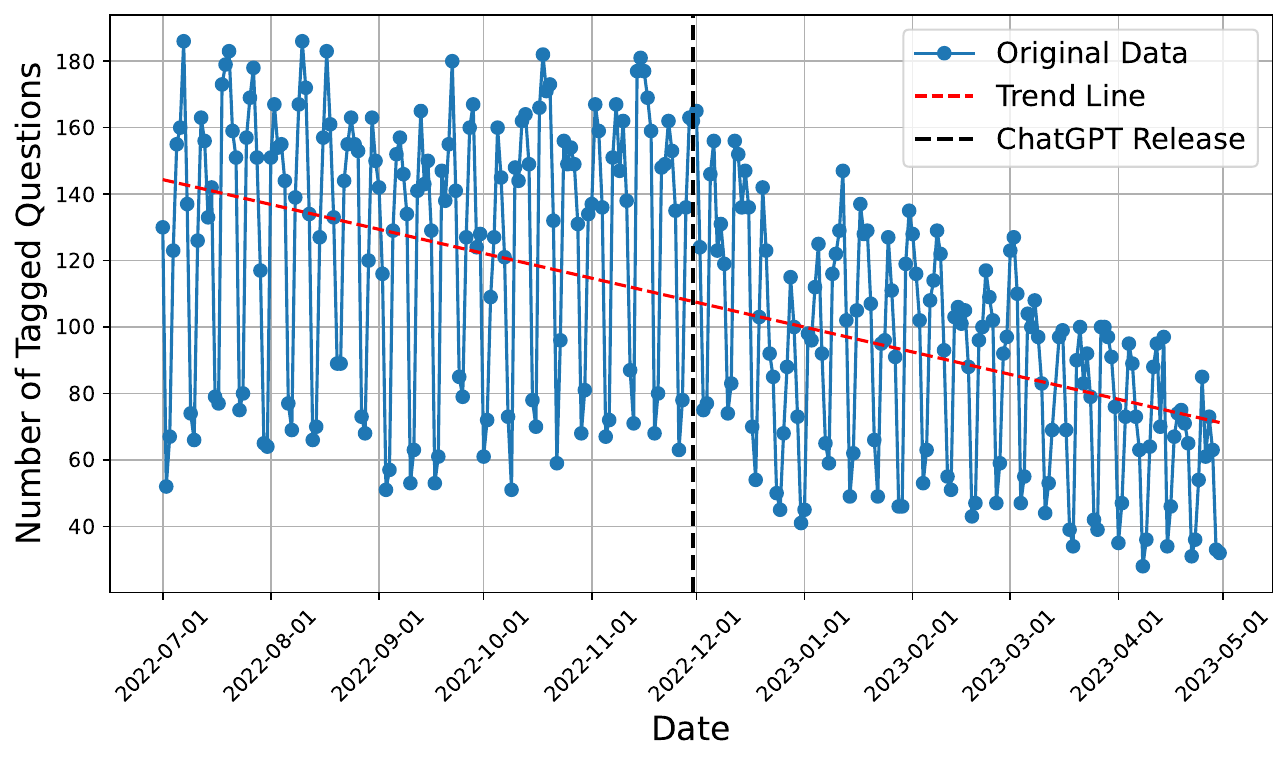}
        \caption{Pandas (Library)}
        \label{fig:trend-pandas}
    \end{subfigure}
    
    \caption{Trend comparison of the number of questions associated with specific tags on \so over time. Here, we just present the results for three tags; in our online appendix, we present the trend analysis for all evaluated tags.}
    \label{fig:trend-analysis}
\end{figure}

\revOne{To better understand the impact of \cgpt's release in different domains, based on the domains that represent a challenge for LLMs (RQ2), we aim to statistically compare the number of questions posted in each domain.}
This way, we explore questions associated with \textit{Programming Languages} and \textit{Frameworks and Libraries}.
\revOne{Figure \ref{fig:trend-analysis} presents the trend analysis for the Top-1 tags of each domain evaluated here. 
Analyzing the charts, we observe that the number of mentions for tags is generally decreasing as time progresses, supporting our previous findings when we observed a general drop in posted content.}
However, we still aim to investigate each domain individually, as they might suffer different impacts.
First, we discuss our findings for programming languages, followed by frameworks and libraries.

\subsubsection*{Consistent Questions' Drop associated with the Top 10 most cited Programming Languages} 
\begin{figure*}
    \centering
    \begin{adjustbox}
    {width=\textwidth,center}
    \includegraphics[width=1\linewidth]{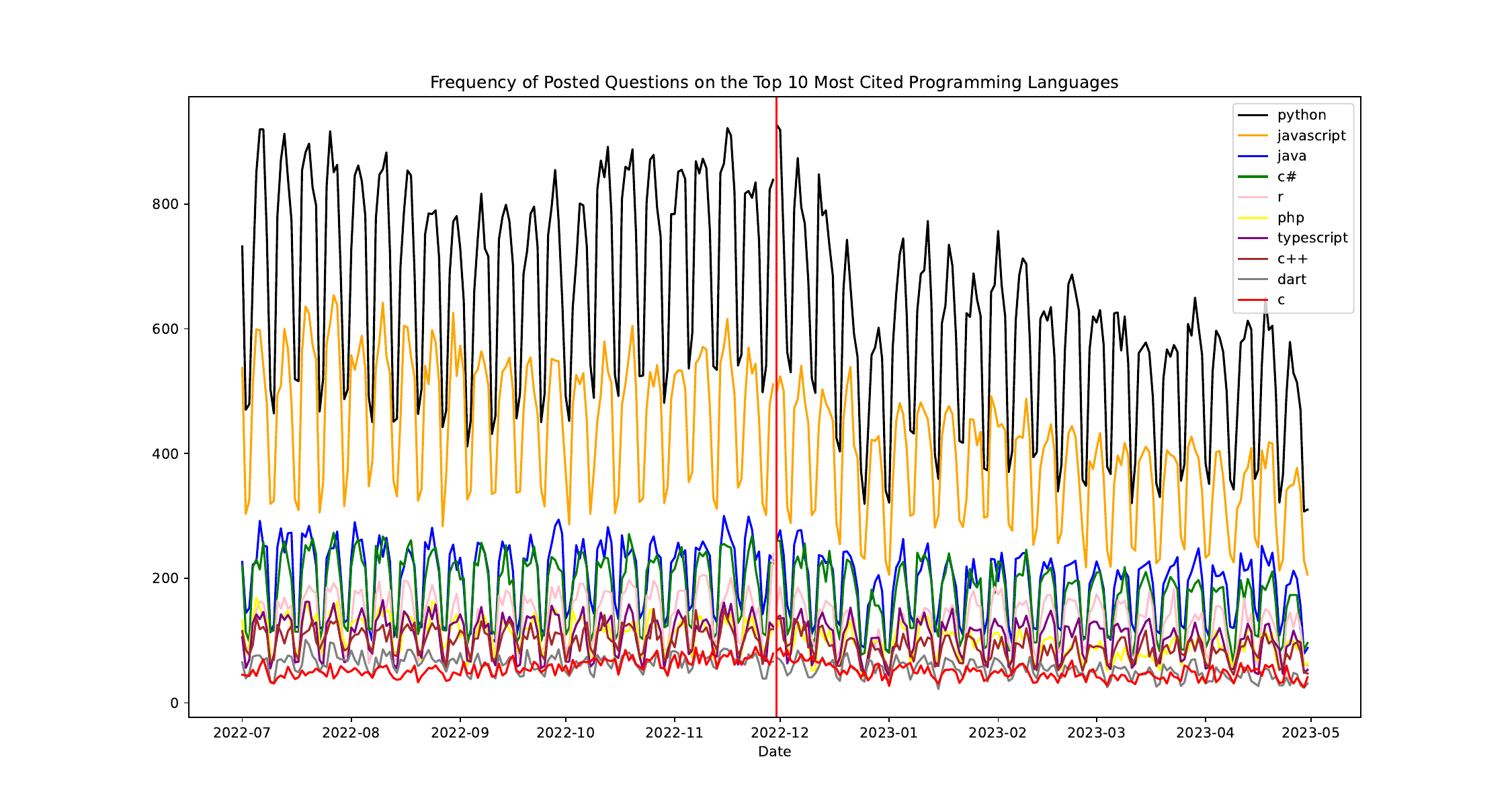}
    \end{adjustbox}
    \caption{Distribution of questions associated with the top ten most cited programming languages on \so questions.}
    \label{fig:chat-languages}
\end{figure*}

\rev{Figure~\ref{fig:chat-languages} shows the distribution of questions associated with the top 10 most cited programming languages.
Overall, we can observe that \texttt{Python} is the most common programming language required by users, followed by \texttt{JavaScript}.
Such results might be motivated by the continuous adoption of these languages for \textit{Machine Learning} and \textit{Web Development}, respectively.
The remaining programming languages present similar frequencies, even oscillating positions with each other, but with a large difference from the top two languages (placed at the bottom of the chart).
}

\rev{
Analyzing the chart, we can observe a clear decline in posted questions after the release of \cgpt for \texttt{Python} and \texttt{JavaScript} (top 2), the same observation made for general posted content analyzed in RQ3. 
However, for the remaining programming languages, there is no clear drop as they present a similar frequency of posted questions before and after the release.
Statistically assessing whether there is a significant difference between the number of posted questions through the \textit{Wilcoxon} rank sum test, we can reject the null hypothesis for all the evaluated programming languages (\textit{p-value} \textless{} 0.01).
}

\revOne{Regarding the effect size, following the previous results reported in this RQ, we observe a moderate effect for most evaluated programming languages (seven out of 10) after the release of \cgpt. However, when checking the effect size after the one-year milestone, we observe a large effect for all evaluated programming languages, revealing a substantial impact on question tagging these programming languages.}

\begin{figure*}
    \centering
    \begin{adjustbox}
    {width=\textwidth,center}
    \includegraphics[width=1\linewidth]{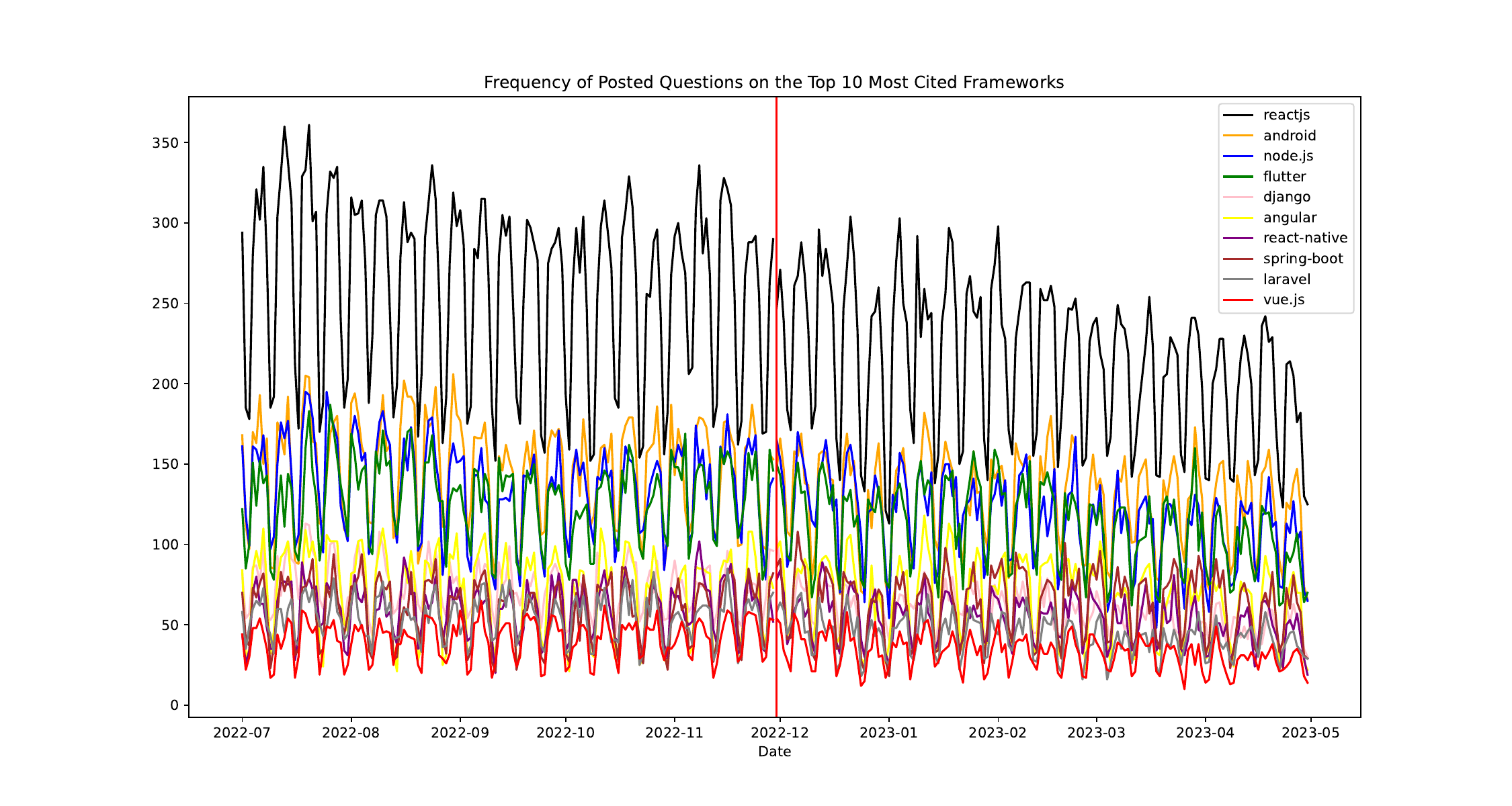}
    \end{adjustbox}
    \caption{Distribution of questions associated with the top ten most cited frameworks on \so questions.}
    \label{fig:chat-frameworks}
\end{figure*}

\begin{figure*}
    \centering
    \begin{adjustbox}
    {width=\textwidth,center}
    \includegraphics[width=1\linewidth]{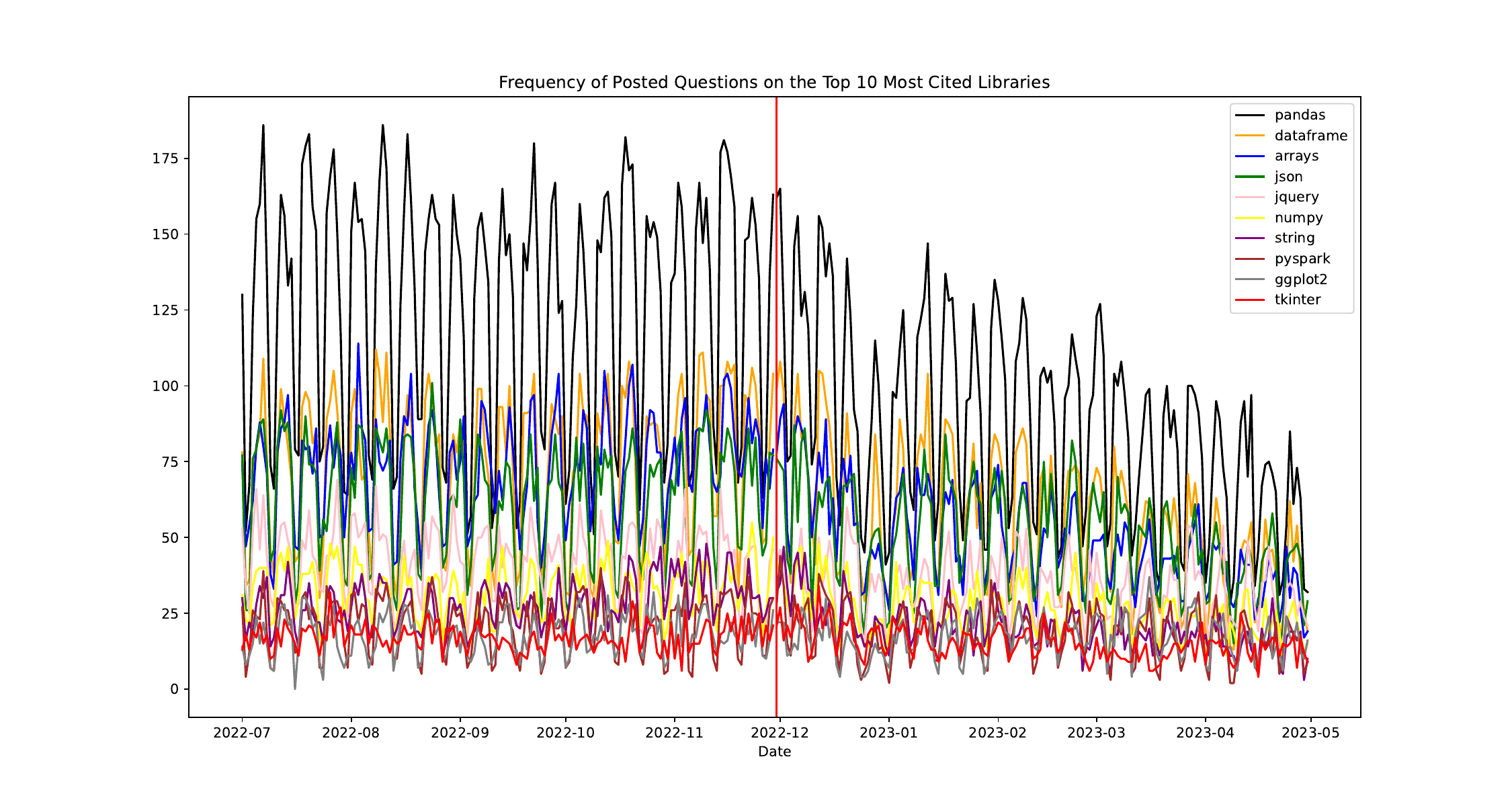}
    \end{adjustbox}
    \caption{Distribution of questions associated with the top ten most cited libraries on \so questions.}
    \label{fig:chat-libraries}
\end{figure*}

\subsubsection*{Different Impact on Most Cited Frameworks and Libraries} 
\rev{Figures~\ref{fig:chat-frameworks} and ~\ref{fig:chat-libraries} show the distribution of questions associated with the top ten most cited frameworks and libraries on \so, respectively. 
Different from the chart for programming languages (Figure~\ref{fig:chat-languages}), we can observe different patterns regarding the distribution of questions here.
Regarding the analysis of \textit{frameworks}, we observe that \texttt{ReactJS} is the most cited framework, largely isolated from the others.
Furthermore, that is the only framework in which we can clearly see a drop in the distribution of questions.
Next, we have \texttt{NodeJS} and \texttt{Android} placed in the second and third positions, respectively.
Although they oscillate in their positions, \texttt{Android} is reported in the second position for most of the evaluated time.
Finally, the remaining frameworks are placed at the bottom of the chart, with some oscillations over time.
\revOne{Different from the programming languages domain, all the frameworks alternate positions except for the first one, showing the ranking of frameworks is more dynamic, with frequent shifts in their relative positions over time.}
Overall, it is also important to highlight the massive presence of frameworks based on \texttt{JavaScript}.
}

\rev{A similar distribution is also observed for the analysis of \textit{libraries}.
As highlighted in Figure~\ref{fig:chat-libraries}, \texttt{pandas} is the most cited library, placed at the top of the chart with a large distance from the others.
As previously discussed for the programming language \textit{Python}, the adoption of \texttt{pandas} might be related to their adoption in \textit{Machine Learning}, as it is commonly used for different tasks, like analyzing, cleaning, exploring, and manipulating data.
Next, we observe a group of four libraries placed in the middle of the chart, while the remaining ones are placed at the bottom, oscillating positions over time.
Finally, similar to the observations made for the \textit{framework} analysis, here we observe a consistent number of libraries supporting different tasks on \textit{Machine Learning} topics, showing the activity of users on these topics.}

\revOne{Aiming to statistically evaluate the impact on the number of posted questions, we also ran the \textit{Wilcoxon} rank sum test.}
As a result, we observe different results when compared to our analysis of \textit{programming languages}.
Regarding the \textit{frameworks}, \revOne{we reject the null hypothesis for nine frameworks (\textit{p-value} \textless 0.05), except for \texttt{Spring Boot} 
(\textit{p-value} = 0.06)}. 
A similar behavior was observed for the \textit{libraries} under analysis, as we reject the null hypothesis for all of them (\textit{p-value} \textless 0.05), except \texttt{ggplot2} (\textit{p-value} = 0.34).
Although a small fraction of subjects did not present a statistical drop in posted questions, \req{when we analyze the posted content after the one-year milestone, we reject the null hypothesis for all evaluated frameworks and libraries (\emph{p-value} \textless 0.01)}. 
\revOne{Regarding the effect size, similar to the previous results for programming languages, we observe a moderate effect for most evaluated libraries and frameworks (six out of 10) after the release of \cgpt. However, when checking the effect size after the one-year milestone, we observe a large effect for all evaluated libraries and frameworks.}

\revOne{On the other hand, we also aim to have an initial insight into domains that exhibited increasing activity over time. 
For that, we first filtered the tags that appeared in all evaluated time windows of this study (before, after, and one year after \cgpt’s release). 
As a result, we identified \num{24564} commonly shared tags. Next, we examined the number of mentions associated with each tag, selecting those with a consistent increase over time, which resulted in 656 tags (2.68\%). 
These tags span various domains, including new versions of programming languages and frameworks, tools, cloud services, and AI/ML-related technologies.
Among these increasing tags, only four exhibited substantial overall activity, each surpassing 1000 mentions: \textit{openai-api}, \textit{vercel}, \textit{powerbi-desktop}, and \textit{next-auth}. 
These technologies have gained significant attention in recent years, with OpenAI API benefiting from the rise of LLMs, and Vercel and NextAuth growing in popularity alongside Next.js.
However, it is important to highlight that the majority of the tags decreased over time, supporting our findings reported here.}

\highlighttwo{
\textbf{RQ3.1 answer:}
Despite the observed general drop in posted content in \so, initially, after the introduction of \cgpt, we observe different impacts on the domains analyzed (\emph{programming languages}, \emph{frameworks} and \emph{libraries}). 
Even analyzing the most cited topics from the previous domains, we observe that, in some cases, there is no statistical decline regarding the volume of posted content. 
However, after the one-year milestone following \cgpt's release, we observe a statistically significant drop in posted content regarding all domains analyzed.
}

\subsection{\revOne{How has the release of \cgpt impacted the activity patterns
of \so users?}}
\label{sec:rq-user-impact}

\rev{In this research question, we further investigate the impact of the release of \cgpt on users' activity on \so.} 
\req{First, we discuss the overall impact on the user's activity, and then, we delve into the challenges faced by users when recurring to LLMs to solve their questions.}

\begin{table*}[]
\begin{tabular}{llcccc}
\hline
\hline
\multicolumn{2}{c}{} & Questioners & Respondents & \begin{tabular}[c]{@{}c@{}}Questioners/\\ Respondents\end{tabular} & Commentators \\ \hline \hline
\multirow{4}{*}{\begin{tabular}[c]{@{}l@{}}Active\\ Users\end{tabular}}         & Pre-Release                                                   & 314,179     & 62,007      & 67,240                                                             & 239,045      \\ \cline{2-6} 
                                                                                & Post-Release                                                  & 293,811     & 56,517      & 57,445                                                             & 206,530      \\ \cline{2-6} 
                                                                                & \begin{tabular}[c]{@{}l@{}}One-Year \\ Milestone\end{tabular} & 174,097     & 31,901      & 30,980                                                             &   116,830           \\ \hline
\multirow{3}{*}{\begin{tabular}[c]{@{}l@{}}Inactive\\ Users\end{tabular}}       & Post-Release                                                  & 259,981     & 49,457      & 56,671                                                             & 179,300      \\ \cline{2-6} 
                                                                                & \begin{tabular}[c]{@{}l@{}}One-year\\ Milestone\end{tabular}  & 259,207     & 48,112      & 50,823                                                             &     171,771         \\ \hline
\multirow{3}{*}{\begin{tabular}[c]{@{}l@{}}Excl.\\ Active\\ Users\end{tabular}} & Post-Release                                                  & 239,613     & 43,967      & 46,876                                                             & 146,785      \\ \cline{2-6} 
                                                                                & \begin{tabular}[c]{@{}l@{}}One-year\\ Milestone\end{tabular}  & 149,503     & 25,284      & 26,489                                                             &     88,257         \\ \hline
\multirow{3}{*}{\begin{tabular}[c]{@{}l@{}}Rem. \\ Users\end{tabular}}           & Post-Release                                                  & 54,198      & 12,550      & 10,569                                                             & 59,745       \\ \cline{2-6} 
                                                                                & \begin{tabular}[c]{@{}l@{}}One-year\\ Milestone\end{tabular}  & 24,594      & 6,617       & 4,491                                                              &      28,573        \\ \hline
\end{tabular}
\caption{User activeness on \so. The three first \emph{Active Users} rows are related to users with active history on pre-, post, and one-year milestone of the \cgpt's release, respectively. The rows for \emph{Inactive}, \emph{Exclusive Active}, and \emph{Reminiscent Users} are calculated by analyzing a specific window along with its preceding windows. For instance, post-release \emph{Inactive Users} are identified as those who were reported as active during the pre-release window.} 
\label{tab:rq-user-frequency}
\end{table*}

\subsection*{Overall Impact on User's Activity Patterns in 
\so}
Table~\ref{tab:rq-user-frequency} presents the general observed frequency of users on \so.
Similar to 
our previous findings, we observe a general drop in active users (2nd and 3rd rows, respectively, in Table \ref{tab:rq-user-frequency}). 
Regardless of the type of user (questioner,  and/or respondent, and commentator), all rates present a consistent drop.
Such observation corroborates the drop in content posted on \so (see Section~\ref{sec:rq1}).
\rev{When we compare the number of users active after the release of \cgpt, based on the different roles analyzed here, we observe that most users were not active before \cgpt's release(4th row in Table \ref{tab:rq-user-frequency}).} 
Some may argue that these new active users could be associated with new accounts created after the release of \cgpt.
However, when checking the creation date of these accounts, most of them were created before \cgpt release (see Table \ref{tab:rq-user-creation-account}).
Since different factors might influence these new active users, like sporadic users, 
it is important to highlight the commitment users have to their community. 
\req{Such motivation might occur due to new technologies that are released and, consequently, further discussions about how to use and advocate for them.}

Although we previously reported a drop in the number of posted questions on \so, these active users posting their new questions contribute to new data being generated and, consequently, new versions of LLMs being trained on them. 
\req{However, keeping users motivated and recruiting new ones to continue the knowledge propagation is a challenge that QA forums must deal with from now on.}
In Section \ref{sec:discussion}, we further discuss the impact and the need for generating data and LLMs.

\subsection*{New Challenges faced by Users in \so}
\req{Once we observe that current active users are new and reminiscent ones, we aim to investigate how these users deal with \so combining with LLMs.}
\rev{For that, we investigate whether users replicate questions on \so, which were previously asked on \cgpt, checking whether the users explicitly mentioned that statement in the question content. 
First, we had to filter the questions asking for clarification about using \cgpt (13\%), like its API (18\% and 7\% for reminiscent and new users, respectively ).
Further exploring the remaining questions, though users mention they have consulted \cgpt for clarifications, 60\% of the questions do not present the answers generated by \cgpt (49\% and 70\% for reminiscent and new users, respectively).
The main reason pointed out by the users is that the answers provided do not fix their problems, as they have tried different proposed suggestions.
For example, when a user asks about \textit{constructing an imbalanced dataset based on a pre-defined gini-coefficient}, they report that the \cgpt could not even generate a working function:\footnote{\url{https://stackoverflow.com/questions/76008808/}}}

\begin{quoting}
    \emph{I have been working on this problem for a few days now, but my programming and math skills are leaking behind. Also, ChatGPT was not able to construct a working function. If someone has any example code or formula I might use, this would be very helpful. Help is really appreciated!!}
\end{quoting}

\rev{For the users that consider the suggestions provided by \cgpt somehow useful, they leveraged 
them as a starting point for further discussions. 
For example, in one question, a user asks for support when \textit{using the Overpass API}.\footnote{\url{https://stackoverflow.com/questions/75558305/}}
Based on the initial solution provided, the respondent analyzed it and provided their suggestion, which, based on the respondent, was adequately addressing the initial issue, while the solution provided by \cgpt was syntactically invalid:}

\begin{quoting}
    \emph{I suggest the following as a starting point, which isn't exactly what you've asked for but way better than what ChatGPT offers as a solution (which is, in fact, invalid Overpass QL syntax).}
\end{quoting}

\rev{In another question, when a user asks about converting \texttt{datetime.datetime} objects in timezone-aware \texttt{datetime} ones, \cgpt provides its answer using the \texttt{pytz} library.\footnote{\url{https://stackoverflow.com/questions/74879121/}}
The respondent answers the question by providing code and using another library (\texttt{pandas}).
Furthermore, the respondent also analyzes the answer provided by \cgpt and warns the questioner about the deprecation of the \texttt{pytz} library.
These scenarios highlight how the discussions started on \cgpt are further extended in \so; while questioners use generated answers to give context to their issues, respondents go further by exploring these answers and providing further feedback.
}

\begin{table*}[]
\centering
\begin{tabular}{lrrrr}
\hline \hline
                      & \multicolumn{1}{c}{Questioners}                             & \multicolumn{1}{c}{Respondents}                              & \multicolumn{1}{c}{\begin{tabular}[c]{@{}c@{}}Questioners \\ and Respondents\end{tabular}}
                      & \multicolumn{1}{c}{Commentators}                              \\ \hline \hline
Accounts \\Pre-ChatGPT  & \begin{tabular}[c]{@{}r@{}}146,115 \\ (61.1 \%)\end{tabular} & \begin{tabular}[c]{@{}r@{}}34,943 \\ (79.59 \%)\end{tabular} & \begin{tabular}[c]{@{}r@{}}38,115 \\ (81.78 \%)\end{tabular} & \begin{tabular}[c]{@{}r@{}}105,043 \\ (71.72 \%)\end{tabular} \\ \hline
Accounts \\Post-ChatGPT & \begin{tabular}[c]{@{}r@{}}93,051 \\ (38.9 \%)\end{tabular}  & \begin{tabular}[c]{@{}r@{}}8,961 \\ (20.41 \%)\end{tabular}  & \begin{tabular}[c]{@{}r@{}}8,495 \\ (18.22 \%)\end{tabular}  & \begin{tabular}[c]{@{}r@{}}41,420 \\ (28.28 \%)\end{tabular}  \\ \hline
                      & 239,166                                                     & 43,904                                                      & 46,610                                                      & 146,463                                                     \\ \hline
\end{tabular}
 \caption{Distribution of users based on account creation date related to \cgpt release date. We only consider users with active accounts and not in anonymous mode, as we could check their account creation date. }
 \label{tab:rq-user-creation-account}
\end{table*}

\rev{
Regarding the topics that represent challenges for \so users, we observe that reminiscent and new users face the same challenges when trying to get answers from \cgpt, and later appeal to \so.
Similar to RQ2 (see Section \ref{sec:rq3}), we observe that questions regarding \textit{Frameworks and Libraries} represent the most recurrent challenge faced by reminiscent and new users (41\% and 38\%, respectively), followed by question targeting \textit{Programming Languages} (22\% and 25\%, for reminiscent and new users, respectively).
Supporting our claims that these topics currently challenge LLMs, users know these limitations as they experience and recognize them in practice.
For example, when a user asks for support regarding starting a new project using multiple frameworks like \texttt{Next.JS}, they consider the provided support of \cgpt was not very helpful. 
They further argue that the failed \cgpt's attempt is due to their limited access to information dated until late 2021:\footnote{\url{https://stackoverflow.com/questions/75861694/}}
}

\begin{quoting}
    \emph{To be honest, the first thing I tried when I ran into walls was ChatGPT 4. It helped a little, but as it says in their documentation, it only knows up to late 2021. All the current libraries have changed [...].}
\end{quoting}

\rev{When conducting our analysis, we observe that more than 20\% of the questions were removed from \so for moderation reasons.
We believe these questions were removed as they violated the politics of \so regarding the adoption of \cgpt to address questions.
Based on our local mined dump, these removed questions have a reference for the adoption of \cgpt by their questioners; however, even when they shared the suggestions, they were used to improve the context of the target question. 
Some may argue that the answers to these questions might have resulted in their exclusion; however, most of these questions even presented associated answers.
In the same way, there was no case of questions properly asking users to validate suggestions generated by \cgpt.
Such observations highlight the complexity of removing questions of \so only based on reference for \cgpt usage.
We further discuss their impact in Section ~\ref{sec:discussion}.}

\highlighttwo{
\textbf{RQ4 answer:}
After the introduction of \cgpt, we observe a drop in the number of active users in \so, even under different roles (questioners, respondents, and commentators). 
Most active users' post-release were inactive before \cgpt, showing that they are reminiscent users, as their accounts were created before the release. 
Finally, reminiscent and new users share the same challenges when addressing their questions on \cgpt and later appealing to \so.
}

\section{Discussion}
\label{sec:discussion}
In this section, we discuss the impact of LLMs on web forums, addressing different challenges from different fields. 

\subsubsection*{LLMs: Impact and Future Direction}
Rio-Chanona et al. \citep{del2023large} discuss the ongoing impact of adopting \cgpt regarding the evolving challenges associated with acquiring the required data for training novel models. 
Our results corroborate this discussion, as we also observe an overall decrease in the posting activities on \so (questions, answers, and comments).
\rev{However, for some specific domains, we did not observe a statistical drop in posting activities, showing that some communities are active and engaged with their mates.
Although different factors might be related to these observations, it is important to understand the motivation behind them.
For example, there is limited support for LLMs for addressing more recent versions of frameworks and libraries.}

\req{Specifically, regarding the impact on the domains analyzed, as reported in Section \ref{sec:impact-domain}, we observed an initial different impact for some libraries and frameworks, later resulting in an overall large impact for all domains analyzed. 
\revOne{Different factors might contribute to this particular result. 
No statistical difference initially observed for specific libraries and frameworks might be caused by the limitations of LLMs (the data on which they were trained).
LLMs do not properly support tasks that involve information generated after their training without further treatment, although they should be able to generalize knowledge. 
Another possible reason is related to the fact that different communities associated with these topics might desire to remain active in supporting their members. 
Even in such cases, though data generation does not represent a challenge, keeping these communities active and supportive might be challenging.
However, the advance of new versions of LLMs, with expanded parameter exploration, demonstrates the potential of LLMs to perform tasks better and, as a result, deliver improved performance.}
}

Nevertheless, despite the declining trend in user-generated content, it is plausible that data generation processes may persist, even through different mechanisms, like the usage of data generated by and through LLMs themselves. 
These alterations would impose specific user requirements to utilize \cgpt for content creation. 
Consequently, data generation may still transpire, although, in altered formats, the critical distinction lies in the potential restriction of its accessibility to the broader public.

\noindent
\subsubsection*{Replacing \so with LLMs}
Our study shows that LLMs perform well when addressing general questions \citep{kashefi2023chatgpt}.
However, for some specific domains, some challenges still need to be improved for the full and broad adoption of LLMs over \so\citep{ray2023chatgpt}. 
Such challenges are recurrent aspects of computer science, based on the constant development of technologies and their replacements over time.
\rev{Trying to overcome this situation, we observe users sharing their issues on \so, after trying to get answers for them on \cgpt.
In these cases, LLMs could be used as a starting point for users, supporting them in properly reporting their issues (context) and providing initial answers.
Later, users could share these initial discussions and even the initial proposed solutions.
Finally, together with their community, they could come to a final solution.
Such behavior was observed for old and new users in \so, showing that this kind of problem is recurrent and requires attention.}

\revOne{For users that opt to rely exclusively on LLMs for addressing their questions, someone may recommend leveraging LLMs 
to accomplish better support for the required tasks.
This way, different techniques can be used, like fine-tuning and merging LLMs.}
\rev{For fine-tuning, the main idea behind this technique is to fine-tune pre-trained models on smaller datasets specific to a particular domain \citep{kasneci2023chatgpt}.
However, fine-tuning models represents an expensive task as it requires time and the same resources used during the pre-training \citep{kaddour2023challenges}. 
This way, fine-tuning models for each new technology or their newer versions is not a feasible approach.}
\so recently announced OverflowAI, a roadmap of practices for the integration of generative AI in the platform.\footnote{\url{https://stackoverflow.blog/2023/07/27/announcing-overflowai/}}
\req{They aim to improve how users interact with their platform and access knowledge. 
For that, they leverage LLMs and generative AI to support users when searching for content, assist them in knowledge creation, and enhance productivity for developers and technical professionals.
Although the main idea is to support users in their daily tasks, and some features are publicly available, to leverage the use of OverflowAI, users might deal with additional costs.
}

\subsubsection*{Prompting LLMs for Tech Questions}
When users ask questions on \so, they are requested to provide the context of their problems \citep{galappaththi2022does}. 
For example, while questions regarding programming languages are more prominent to have associated code with and even documentation, questions exploring GUI and setup of tools/environment might adopt non-textual information \citep{nasehi2012makes}.
Although these non-textual sources provide details that could be hard and time-consuming for users to provide with textual elements, they reflect a limitation for current LLMs (no-textual element support).
To overcome this limitation, users must adapt how they prompt LLMs (detailed textual information) \citep{white2023prompt}, while LLMs could also further explore new ways of supporting non-textual information sources \citep{li2023blip}.
\req{In the same way, we observe that when LLMs were prompted and given code snippets associated with supplementary data, like error messages and outcomes, they reached higher similarity when compared to the accepted answers. For coding tasks like this, we encourage users to adopt such an approach, as that supplementary data adds important context information that could guide the LLM when addressing the questions.}

\req{Our findings show that LLMs perform well when addressing \so questions, highlighting their capabilities for programming language tasks.
As previously mentioned, supplementary data could be used as input with code snippets.
For questions from other domains, there is a gap regarding the supplementary data that could support LLMs.
Researchers could conduct further studies to discover and provide these new guidelines.
Based on these guidelines, users could better structure their questions and find better support from LLMs. 
In the same way, we believe further investigation could be performed to explore LLMs addressing unanswered questions in \so.
For that, the previously mentioned guidelines could be used to also structure the questions given to LLMs.
}

\so users also provide external sources of information when asking or answering questions.
For the answers with high textual similarity, 33\% of them have associated external information, while 70\% provide links for supporting official documentation \citep{baltes2020contextual}, which is beneficial to the user to double-check the solution or learn more about the problem.
LLMs did not report any external link or material when generating their answers, which could be understood as documentation or reference for users.
Providing such information might represent a valuable option for the users to reflect on their trust in the provided solution, or even educate them to further investigate the problem by themselves \citep{uddin2019understanding, robillard2011field}.

\subsubsection*{LLMs Impact on Education}
Our results show that LLMs perform well when addressing questions related to generic problems and programming languages.
\revOne{Although users can benefit from using LLMs during programming tasks, they are expected to be able to critically analyze the generated code and, if necessary, make adjustments to improve it \citep{marsicano2017team}}.
For example, when discussing an answer provided by \llama in Section \ref{sec:rq2}, we observe the proposed solution does not fix the problem.
Although the code could be easily fixed, that is a decision that requires the user to understand the code, locate the bug, and then, apply the required changes \citep{johnson2019empirical, oliveira2020evaluating}.
\req{Another scenario takes place when LLMs suggest code using deprecated libraries, that could potentially introduce vulnerabilities in the code \citep{decan2018impact} (see Section \ref{sec:rq-user-impact}). 
This case requires more attention, as the user must be mature enough to evaluate the impact of handling these vulnerabilities.}
These issues could be addressed by exploring new features on LLMs, like exploring conventional programming tools to detect general errors or vulnerabilities.
Recently, OpenAI released a new feature, Code Interpreter, which allows the \cgpt to execute Python code \citep{codeInterpreter}.

\revOne{The same discussion also occurs when students benefit from these models, by prompting LLMs for different purposes, specifically on programming tasks \citep{surameery2023use}}.
Given the power of these models, students should use them carefully, not to limit them in their learning process.
\rev{For example, when answering RQ4, we observe some users posting answers generated by \cgpt in their questions and explaining the reasons why the generated answers did not resolve their issues.
Although this critical thinking is an expected skill for students, they might develop such skills with time through proper learning.}
Besides the common skills regarding programming, students must also develop new skills to improve their experiences when prompting the LLMs, like prompt engineering and related fields \citep{strobelt2022interactive, white2023prompt}.

\subsection{Implications}

\revOne{Our findings offer valuable insights into the practical impact of using LLMs as assistant tools in software development. 
Additionally, they highlight the impact of replacing QA forums with LLMs and the broader implications for future generations of models. 
Next, we discuss how our results affect different stakeholders.}

\revOne{\textbf{Implications for Developers:} Overall, we believe that \so users will continue using LLMs as assistant tools. 
While these models can generate useful responses, their limitations in certain domains, as reported here, may lead to incorrect conclusions, emphasizing the need for careful validation. 
While Stack Overflow users are accustomed to reading entire threads and multiple answers \citep{zhang2019reading}, LLM users typically receive a direct response without the need for extensive browsing. 
However, they can engage in a back-and-forth discussion with the model, simulating a one-on-one interaction and challenging its responses.
}

\revOne{\textbf{Implications for Communities in \so:} With the decline of posting activity in \so, communities should foster discussions, encourage peer-reviewed responses, keep users motivated, and actively contribute to knowledge sharing. 
Knowing that the role of these communities is not restricted to Q\&A forums but also extends to maintaining high-quality technical discussions, curating reliable information, and mentoring new contributors, it is crucial to reinforce engagement and sustain a collaborative environment. 
By doing so, these communities can complement LLM-generated responses, ensuring that software developers continue to have access to accurate and well-reviewed knowledge.}

\revOne{\textbf{Implications for Model Maintainers/Researchers:} As previously discussed, the decline in high-quality data generation will impact the development of new models. 
Knowing that domain-specific data is essential for improving future models and ensuring equitable access to reliable information, researchers and model maintainers should invest in strategies to sustain and expand high-quality datasets. 
We believe different strategies can be adopted, like fostering collaborations with developer communities to curate and validate domain-specific knowledge, and encouraging open data-sharing initiatives. 
Additionally, hybrid approaches that integrate human expertise with LLM-generated content can help maintain knowledge quality and relevance over time.
}

\revOne{Understanding these challenges is crucial for both practitioners and researchers seeking to enhance AI-assisted programming. Additionally, our study highlights the role of developer communities in complementing LLM-generated knowledge, reinforcing the importance of peer-driven support. Finally, we discuss the significance of data availability in improving future models and ensuring equitable access to high-quality training data}

\section{Threats to Validity}
\label{sec:threats}
Our study design introduces particular validity concerns, which we discussed as follows.

\textit{\textbf{Construct to Validity:}} 
While we assess the potential of Language Model Models (LLMs) to address questions posted on Stack Overflow, our analysis is limited to determining the similarity between the answers generated by LLMs and the answers labeled as \texttt{accepted} by \so users. 
We decided to compare with the accepted answers as they truly align with the specific inquiries posed by the original asker, ensuring a closer correspondence to the requester's intent.
It is plausible that the answers generated by LLMs could effectively address the underlying questions, but 
deviate from the answers marked as accepted, leading to potential misclassification.
Even when comparing the generated answers with the entirety of answers available for a given question, there remains a possibility that the generated responses differ from these answers but still offer valuable solutions.

\revOne{When randomly selecting the group of questions used to answer RQ1 and RQ2, we did not filter out questions that referred \cgpt or \llama. 
Knowing that this could introduce bias by including questions that directly reference these models, we verified its impact by checking occurrences of these terms and re-running our analysis. Since only two occurrences were found (out of 384 questions) and the results remained unchanged, we believe this issue does not compromise the validity of our findings.}

While our data collection includes information from \so that postdates the inception of the training data used for ChatGPT, it remains uncertain whether the \llama model was also exposed to this more recent data during its training. 
Consequently, there exists a potential for the outcomes presented herein by \llama to be influenced by memorization effects~\citep{carlini2022quantifying}, which may introduce a degree of bias into our results.
\revOne{To address this concern, we could apply the same approach used for ChatGPT by filtering the data based on LLaMA’s knowledge cutoff. Specifically, LLaMA’s initial cutoff in September 2022 suggests that potential bias may have been introduced when collecting data between July and November 2022 (ChatGPT’s release). Furthermore, subsequent tuning updates, including those up to July 2023, may have influenced LLaMA’s responses, potentially altering its behavior. 
However, despite this potential threat, our findings show that both models exhibit similar trends, with ChatGPT outperforming LLaMA. 
This consistency suggests that any memorization effects are unlikely to meaningfully impact our overall conclusions.}

\textit{\textbf{Internal Validity:}}
Since we compute the textual similarity using the cosine metric, the potential for misclassification of answers exists due to structural and stylistic disparities relative to the accepted answers.
As previously discussed, that metric might lead to false negatives in our analysis.
To mitigate this concern, we have incorporated an additional dimension to our analysis by assessing similarity through semantic evaluation by prompting LLMs.
We know this method also introduces certain potential limitations, which we endeavor to mitigate by employing a specified similarity scale to standardize and refine semantic similarity assessment.
Concerning sentiment analysis, it is noteworthy that the tone and sentiment inherent to the input provided in the prompt can influence the tone assumed by the LLM; however, we prompt each LLM with the same input, trying to eliminate possible associated bias.

When investigating the challenges faced by LLMs to address the questions, we rely on performing manual analysis.
To mitigate potential biases, an initial exploration was performed in order to establish the information to be extracted, resulting in a spreadsheet adopted as a guide to the researcher during this step. 
That spreadsheet encompasses a comprehensive list of labels and their potential values, serving as a guiding reference for the researcher throughout the analytical process, ensuring systematic and rigorous scrutiny.
\revOne{Someone may argue that we could have missed some types of information; however, by verifying the type of information that can be shared when asking a question on \so, we can ensure that our analysis covers 
all the types currently supported.\footnote{\url{https://stackoverflow.com/editing-help}}}

\revOne{When evaluating the impact of activity patterns on users in \so, our goal was to capture the behavioral impact directly associated with ChatGPT’s release. 
Including data from a more distant pre-ChatGPT period could introduce noise from unrelated historical factors (e.g., long-term platform trends, policy changes, or seasonal patterns), making it harder to isolate ChatGPT’s effect. By restricting the analysis to a short pre-release window, we aim to reduce the risk of confounding effects and maintain focus on the most relevant comparison.
However, we acknowledge that further studies are required to understand and measure possible co-founding factors associated with ChatGPT's release.}

\textit{\textbf{\revOne{Reliability Validity:}}} \revOne{For performing our semantic similarity analysis, two different LLMs were used; both of which led to similar conclusions, showing an overall disagreement with our manual analysis.
However, when running the same analysis multiple times with \cgpt, we also observed a low inter-agreement (Krippendorff alpha = 0.34), indicating some level of inconsistency in its responses. 
This suggests that while another human annotator might introduce further precision, the fundamental conclusions are unlikely to change, given the weak agreement between LLM runs.} 

\textit{\textbf{\revOne{External Validity:}}}
Our results are limited in the context of technical questions posted on \so and the LLMs evaluated here.
Although we can not generalize our findings to other web forum communities, the concerns we have elucidated may have broader applicability to web forums in a more general context.
Regarding the LLMs evaluated here, we aimed to explore how different LLMs perform based on our study setup.
However, future versions of the same LLMs we evaluate, and other new proposed ones, might unveil nuances and aspects beyond the scope of our current investigation.

\section{Related Work}
\label{sec:related-work}
LLMs have been explored for different purposes, especially after the release of \cgpt.
In this section, we discuss some related studies, highlighting the differences reported by our study.
\revOne{First, our closest related studies usually investigate correctness, while we investigate reliability, since technically correctness may still be misleading
or harmful to users in some contexts.}
Second, overall, most related studies focused on exploring \so under the exclusive perspective of \cgpt.
In this work, we take a different turn by also evaluating \llama and further comparing its results with \cgpt.
In the same way, while related work mostly relies on prompting LLMs for specific domains of questions, we adopt a systematic approach to broadly select random questions from different domains and time spaces.
This way, besides evaluating the reliability of generated questions, we also explore potential bias resulting from previous training data and the domains in which LLMs struggle to provide support.

Widjojo and Treude \citep{widjojo2023addressing} investigate how LLMs can support developers when fixing code errors.
To that end, the authors examine 100 code snippets with compiler errors from previous work and new ones generated by them.
Next, they explore different combinations of configurations to prompt \so and \cgpt (versions 3.5 and 4) to address each error.
The authors report that when prompting \cgpt with the offending code snippet, the LLM reports contents more instrumental and helpful.
Furthermore, they observe \cgpt 4.0 outperforms its previous version (3.5), as it considers the context associated with the error, while \cgpt 3.5 focuses on straightforward solutions.

Similar to our goal, Kabir et al. \citep{kabir2023answers} focus on comparing the answers provided by humans and \cgpt to \so questions.
They select a sample of 517 questions and prompt \cgpt to answer them.
The authors investigate the correctness of \cgpt answers by adopting a manual analysis.
They report that 52\% of generated answers are incorrect, while 22\% are consistent with human answers.
Despite the different processes to define their sample and the analysis performed, and based on the computed cosine similarity, our results show better performance of \cgpt as around 85\% of the generated answers presented a good similarity (higher than 0.5).

\rev{In the same way, Delile et al. \citep{delile2023evaluating} investigate the usage of \cgpt-3.5 to address privacy-related questions from \so.
The authors adopt a different approach by analyzing 82 out of 932 mined questions from a broader time window (2016 to 2023), which could impact their results due to using related training data.
Next, they manually compare the generated answers by the LLM with the accepted ones. 
The authors report that the LLM presented itself as an alternative solution for addressing the \so questions. 
Overall, although the results conform with ours, they do not explore the topics that challenge LLMs.
Furthermore, the study has some threats we try to address here, like the memorization \citep{carlini2022quantifying} and diversity of evaluated LLMs.}

\revOne{Similarly, Oishwee et al. \citep{oishwee2024large} investigate \cgpt's ability to answer \so questions related to Android permissions. 
Like us, the authors mine \so questions, prompt them to the LLM, and manually evaluate how closely the generated responses match the accepted answers.
Since their dataset overlaps with \cgpt-3.5’s training cutoff, potential biases may have influenced the results. 
The authors report that 53.26\% of the generated answers align with accepted Stack Overflow answers, while an additional 27\% align with positively voted responses when they do not match the accepted ones, showing the potential of \cgpt to assist developers in their daily tasks.}

Pinto et al. \citep{pinto2023large} investigate the usage of \cgpt to provide feedback to developers.
They selected six open-ended questions from two subjects (\dcircle{1} caching, and \dcircle{2} stress and performance testing), that were asked for one expert in each subject and a group of 40 developers (questionnaire).
Unlike our work, which prompts \cgpt to answer open-ended questions, the authors prompt the LLM to grade the answers provided by the participants. 
Additionally, when grading the experts' answers, the authors prompt \cgpt to correct their answers, resulting in a general consensus among experts concerning the quality and accuracy of the explanations provided.

Liu et al. \citep{liu2023better} investigate the potential of adopting \cgpt as a code assistant tool for developers.
For that, they conducted a study with a group of students, exploring questions related to \textit{general algorithms}, \textit{library}, and \textit{debugging}.
This way, the participants were asked to address the selected question, consulting \cgpt and \so.
The results show that adopting \cgpt for answering the \textit{algorithm} \textit{library} questions outperforms \so regarding participants providing correct answers, supporting our findings here.

\rev{Previous studies also investigate the impact of \cgpt on posting activities on Q\&A communities. 
While Xue et al. \citep{xue2023can} focus on the impact of posting questions exclusively in \so, Rio-Chanona et al. \citep{del2023large} adopt a different approach by considering other popular web forum platforms.
While the first study focuses on posted questions, the last explores general posting (questions and answers).
Ultimately, both studies collect posting information from all evaluated platforms and propose a model to estimate the effect.
Similar to our findings, Rio-Chanona et al. \citep{del2023large} observe a significant decrease of 15.6\% in the posting activities on \so when compared to the other platforms, while Xue et al. \citep{xue2023can} report a significantly negative reduction of the posting by 2.64\%.
Furthermore, they also report a decrease in the readability of posted questions and cognition level (2.55\% and 0.4\%, respectively).
\req{We go one step further by investigating not only the immediate impact after the release of \cgpt, but also the impact after one year of its release. 
Furthermore, we also investigate the impact on specific domains in \so, reporting that no statistically significant impact was initially observed for some domains. However, after one year of the release, the decline was constant and significant for all domains evaluated.
Similarly, Burtch et al. \citep{burtch2023consequences} also investigate that impact, but not grouping related topics as we do here.}
}

\rev{Similar to our study, Burtch et al. \citep{burtch2023consequences} and Xue et al. \citep{xue2023can} also investigate the impact of \cgpt on \so users, focusing on the access traffic and the way users deal with Stack Overflow, respectively. 
Xue et al. \citep{xue2023can} report that new users are negatively impacted by \cgpt by asking longer and less readable questions.
We take a different perspective by investigating the drop in users' participation in their roles in \so (questioner, respondent, and commentator), and the new types of users.
Furthermore, we explore the challenges current users face regarding the adoption of \cgpt and \so to address their questions.}

\section{Conclusion}
\label{sec:conclusion}
With the release of different LLMs, especially those targeting different tasks in software engineering, practitioners and researchers have investigated means for ensuring their proper usage.
Large Language Models such as \cgpt and \llama are known for their potential to support users in their work, commonly treated as assistant tools.
In this work, we investigate the potential of using LLMs to address \so questions.
We have conducted an empirical study assessing the reliability of the answers generated by these LLMs for existing questions in \so while identifying possible challenges.

Overall, our results report that answers generated by \cgpt and \llama-2 show a high degree of textual similarity to the ones accepted in \so.
Although \cgpt outperforms \llama regarding the textual similarity of generated answers, \llama performs well, representing a good, strong, and free option for the general audience.
While LLMs perform well on questions associated with general and specific problems on \textit{programming languages}, they face some challenges when addressing \textit{frameworks and libraries} questions.
\rev{As a result, users might appeal to \so when they do not get the expected support for their issues on LLMs.}
Finally, we also observe a significant decline in user activity on \so since the release of \cgpt (questions, answers, and comments). 
\rev{Initially, for some domains, there is no statistically significant difference regarding the frequency of posted questions, showing that their communities keep active on \so.
However, after one year of the release, we observed a constant and significant decline for all domains evaluated.}

These findings reinforce the capability of LLMs to perform different tasks, shedding light on discussions regarding their impact.
Replacing tools, like \so with LLMs, represents a huge change that is not prudent to take for now.
Currently, LLMs still have to further explore new features to provide generalizable, reliable, and valuable knowledge for their audience, while users are required to develop new skills to improve their experiences when adopting these models.

\section{Acknowledgements}
\label {acknowledgements}
We thank Arghavan Dakhel for her support in setting up the environment for this study. We also thank the anonymous reviewers for their valuable comments on improving an earlier version of this paper. This work is funded by the following organizations and companies: Fonds de Recherche du Quebec (FRQ), Natural Sciences and Engineering Research Council of Canada (NSERC), and the Canadian Institute for Advanced Research (CIFAR). However, the findings and opinions expressed in this paper are those of the authors and do not necessarily represent or reflect those organizations/companies.

\section*{Declaration}
\label{sec:data_availability}
\textbf{Conflicts of Interest}: The authors declare that they have no known competing financial interests or personal relationships that could have appeared to influence the work reported in this paper.

\noindent
\textbf{Data Availability:} To promote open science and facilitate reproducibility, we make all our artifacts available to the community.
This includes the datasets used in our experimentations, the source code for the scripts that were used, the results produced, and any other artifacts related to our study: \url{https://github.com/leusonmario/chat-stack}, \url{https://doi.org/10.5281/zenodo.15086541}.

\bibliographystyle{elsarticle-harv}

\end{document}